\documentstyle[aps,pra]{revtex}

\begin{document}
\draft
\tighten

\title{Collisionless dynamics of dilute Bose gases:\\ 
Role of quantum and thermal fluctuations}

\author{ S. Giorgini}

\address{Dipartimento di Fisica, Universit\`a di Trento,
\protect \\
and Istituto Nazionale di Fisica della Materia, I-38050 Povo, Italy} 

\maketitle

\begin{abstract}

{\it We study the low-energy collective oscillations of a dilute Bose gas at 
finite temperature in the collisionless regime. By using a time-dependent 
mean-field scheme we derive for the dynamics of the condensate and noncondensate 
components a set of coupled equations, which we solve perturbatively to second 
order in the interaction coupling constant. This approach is equivalent to the 
finite-temperature extension of the Beliaev approximation and includes corrections 
to the Gross-Pitaevskii theory due both to quantum and thermal fluctuations. For a 
homogeneous system we explicitly calculate the temperature dependence of the 
velocity of propagation and damping rate of zero sound. In the case of harmonically
trapped systems in the thermodynamic limit, we calculate, as a function of 
temperature, the frequency shift of the low-energy compressional and surface modes.} 

\end{abstract}


\section{Introduction}

The study of collective excitations is one of the main areas of interest
for the experimental and theoretical research activity in trapped Bose-condensed
gases (for a review of experimental and theoretical investigations see 
respectively \cite{SS} and \cite{DGPS99}). At low temperatures, the frequencies of
the low-energy collective oscillations of the condensate have been measured with 
great accuracy \cite{JILA96,MIT99}, and found in very good agreement with the predictions 
of the mean-field Gross-Pitaevskii theory \cite{ERBDC96,S96}. In a series of experiments
carried out at JILA \cite{JILA97} and MIT \cite{MIT98} the excitations of a trapped Bose gas
have also been explored as a function of temperature. The main features are: on the one hand 
oscillations of both the condensate and the thermal cloud are visible and, on the other hand,
the oscillations are increasingly damped as temperature is raised and temperature dependent
frequency shifts are also observed.
A theoretical description which correctly accounts for these phenomena has not yet been fully 
developed.

At finite temperature the dynamics of Bose-condensed systems is complicated. Depending on the 
temperature, density and frequency of the excitations one is probing different regimes (for an 
exhaustive discussion see the books \cite{NP90} and \cite{G93}). If the frequency $\omega$ is 
much smaller than the inverse of the typical collision time $\tau_c$: $\omega\tau_c\ll 1$, the 
excitations are collective collisional modes, which are described by the theory of two-fluid 
hydrodynamics. In terms of length scales this regime is equivalently defined by the condition: 
$\lambda_{ex}\gg \ell_{mfp}$, where $\lambda_{ex}$ is the wavelength of the excitation and 
$\ell_{mfp}$ is its mean free path. At low temperatures and low values of the density the mean 
free path becomes comparable with the size of the system. In this case, which corresponds to the 
condition $\omega\tau_c\gg 1$, one is probing the collisionless regime, which is properly described 
by mean-field theories. 
Collisionless modes can be further distinguished into collective and single-particle excitations, 
depending on whether the excitation energy lies respectively well below or above the chemical 
potential $\mu$. Single-particle excitations have wavelength much smaller than the healing length 
of the condensate, which is defined as $\xi=1/\sqrt{8\pi a n_0}$, where $a$ is the $s$-wave scattering 
length and $n_0$ is the condensate density. On the contrary, collective modes satisfy the condition: 
$\lambda_{ex}\gg\xi$. Finally, in harmonically trapped systems, collective modes can behave 
semiclassically if their energy is much larger than the typical trapping energy: $\hbar\omega_{ho}
\ll\hbar\omega\ll\mu$, where $\omega_{ho}$ is the harmonic oscillator frequency. If instead 
$\hbar\omega\sim\hbar\omega_{ho}$, the discretization of levels becomes important and one is not 
allowed to treat the excitation as quasiclassical.         

Collective modes in the collision-dominated regime have been investigated in harmonically trapped
systems by several authors \cite{HYD}. The present work is focused on the study of collective 
excitations in the collisionless regime. In the last years a large number of theoretical papers
have appeared in the literature addressing this problem. Mean-field approaches, which extend to 
finite temperature the Gross-Pitaevskii equation for the order parameter, have been put forward 
\cite{G96} and applied to the calculation of the low-energy modes in traps 
\cite{HZG97,HDB98,SZ99}. However, in these approaches the noncondensate component is treated as a 
static thermal bath and its dynamic coupling to the oscillations of the condensate is neglected. 
The results obtained for the collective modes do not adequately reproduce the features observed 
in experiments, in particular these approaches do not account for the damping of the excitations. 

More accurate time-dependent mean-field schemes have been proposed \cite{PB96,MT97,G98,RB99}, 
which describe
the coupled dynamics of the condensate and noncondensate components. These methods have been 
applied to the study of damping in trapped systems \cite{G98} and agree with results obtained 
from perturbation theory \cite{PS97,FSW98}. Explicit calculation of the damping rate of the 
low-energy modes in harmonic traps has been carried out in \cite{FSW98,BS98,GP99,RCGS99} and found in 
good agreement with experiments. Similar methods have also been applied to the calculation of 
the temperature dependence of the frequency shifts in the collisionless regime.            
Bijlsma and Stoof \cite{BS98-1} have used a variational ansatz to describe the time evolution of the 
condensate and the thermal cloud and have calculated the frequencies of the coupled modes in which 
the two
components move either in phase or out of phase. These authors also suggest that the avoided crossing 
between the in and out of phase modes might be the reason of the features observed for the $m=0$ 
mode at JILA \cite{JILA97}. Olshanii has explicitly analyzed the JILA $m=2$ mode \cite{JILA97}, 
suggesting that the observed temperature dependence of the excitation frequency might be due to a
strong resonance between the oscillation frequency of the condensate and one of the eigenfrequencies
of the thermal cloud \cite{O98}. Fedichev and Shlyapnikov \cite{FS98} have developed a 
Green's function perturbation scheme for inhomogeneous Bose-condensed gases at finite temperature
and have calculated energy shifts and damping rates of quasiclassical collective modes, which 
satisfy the condition $\hbar\omega_{ho}\ll\epsilon\ll\mu$, being $\epsilon$ the energy of the 
excitation. Very recently, Reidl {\it et al.} \cite{RCGS99} have calculated by the dielectric 
formalism the frequency shift of the $m=0$ and $m=2$ modes, and compared the results with the 
JILA experiments.  
  
In the present work we derive, within a time-dependent mean-field scheme, coupled equations
for the dynamics of the condensate and noncondensate components. These equations are solved 
perturbatively to 
second order in the interaction coupling constant. For homogeneous systems this approach is equivalent
to the finite-temperature extension of the Beliaev approximation discussed in \cite{SG98}. 
In the homogeneous case we give explicit results for the temperature dependence of the velocity of
zero sound, which include effects beyond the Bogoliubov theory. We also apply our analysis to 
harmonically trapped systems in the thermodynamic limit. In this regime, which is reached for systems
with a very large number of trapped particles, one can use the Thomas-Fermi approximation for the 
condensate and neglect finite-size effects. Under these conditions, which are not difficult to 
realize in experiments (see e.g. \cite{MIT98}), the frequencies of the collective modes are found
to change with temperature due to static and dynamic correlations beyond the Gross-Pitaevskii theory.   
We calculate, as a function of temperature, the frequency shifts of the lowest compressional and 
surface 
modes. We find that at the intermediate temperatures $T\sim$ 0.6-0.7 $T_c$, where $T_c$ is the 
transition
temperature, the fractional shift due to beyond mean-field effects is of the order of 5\%. 
This result should be compared with the corresponding correction   
predicted at very low temperatures \cite{PS98,BP99} and arising from quantum fluctuations, which 
turns out to be typically of the order of 0.5\%. 

The structure of the paper is as follows. In Sec. II we develop the time-dependent mean-field 
scheme and derive coupled equations for the small-amplitude oscillations of the condensate and
noncondensate components. Sec. III is devoted to spatially homogeneous systems. First we develop 
the perturbation scheme and hence we calculate to second order in the interaction coupling 
constant the equation of state of the system and the speed and damping rate of
zero sound. In Sec. IV we apply the same perturbation scheme to harmonically trapped systems in the 
thermodynamic limit. We calculate the temperature dependence of the frequency shift of the low-energy
collective modes and discuss the comparison with experiments. Finally, we show that at zero 
temperature our approach reproduces the hydrodynamic equations of superfluids.   

\section{Time-dependent mean-field scheme}

Our starting point is the grand-canonical Hamiltonian of the system in the 
presence of an inhomogeneous external potential $V_{ext}({\bf r})$. 
In terms of the creation and annihilation particle field operators 
$\psi^{\dagger}({\bf r},t)$ and $\psi({\bf r},t)$, the Hamiltonian
takes the form
\begin{eqnarray}
H^{\prime} \equiv H-\mu N &=& \int d{\bf r} \;\psi^{\dagger}({\bf r},t)
\left( -\frac{\hbar^2\nabla^2}
{2m} + V_{ext}({\bf r}) - \mu \right) \psi({\bf r},t) \nonumber \\
&+& \frac{{\rm g}}{2} \int\;d{\bf r} \;\psi^{\dagger}({\bf r},t)
\psi^{\dagger}({\bf r},t)
\psi({\bf r},t)\psi({\bf r},t) \;\;.
\label{gch}
\end{eqnarray}
In the above equation we have assumed a point-like interaction between 
particles $V({\bf r}-{\bf r}')={\rm g}\delta({\bf r}-{\bf r}')$, 
with the coupling constant ${\rm g}$ given by the expression 
${\rm g}=4\pi\hbar^2a/m$, to lowest order in the $s$-wave scattering 
length $a$. The equation of motion for the particle field operator 
follows directly from the Heisenberg equation and reads
\begin{eqnarray}
i\hbar\frac{\partial}{\partial t}\psi({\bf r},t) &=& \left[\psi({\bf r},t),
H^{\prime}\right] \nonumber \\
&=& \left(-\frac{\hbar^2\nabla^2}{2m}+V_{ext}({\bf r})-\mu\right)
\psi({\bf r},t) + 
{\rm g} \, \psi^{\dagger}({\bf r},t)\psi({\bf r},t)\psi({\bf r},t) \;\;.
\label{heq}
\end{eqnarray}

The dynamic equations derived in this section correspond to the linearized 
time-dependent Hartree-Fock-Bogoliubov (TDHFB) approximation. This 
self-consistent mean-field scheme is based on the following prescriptions 
(we use the notations of Ref. \cite{G96}):
\begin{eqnarray}
a)\;\;\; && \psi({\bf r},t)=\Phi({\bf r},t)+\tilde{\psi}({\bf r},t) 
\nonumber \\
               && \Phi({\bf r},t)=\langle\psi({\bf r},t)\rangle \nonumber \\
               && \langle\tilde{\psi}({\bf r},t)\rangle=0 \nonumber \\
&& \nonumber \\
b)\;\;\; && \langle\tilde{\psi}^{\dagger}({\bf r},t)\tilde{\psi}
({\bf r},t)\rangle = \tilde{n}({\bf r},t) \nonumber \\
               && \langle\tilde{\psi}({\bf r},t)\tilde{\psi}({\bf r},t) 
\rangle = \tilde{m}({\bf r},t) \nonumber \\
&& \nonumber \\
c)\;\;\; && \tilde{\psi}^{\dagger}({\bf r},t)
\tilde{\psi}^{\dagger}({\bf r},t)\tilde{\psi}({\bf r},t)
\tilde{\psi}({\bf r},t)= 4\tilde{n}({\bf r},t)\tilde{\psi}^{\dagger}({\bf r},t)
\tilde{\psi}({\bf r},t) \nonumber \\
               && \;\;\;\;\;\;\;\; + \tilde{m}({\bf r},t)\tilde{\psi}^{\dagger}({\bf r},t)
\tilde{\psi}^{\dagger}({\bf r},t) + \tilde{m}^{\ast}({\bf r},t)\tilde{\psi}({\bf r},t)
\tilde{\psi}({\bf r},t)
\nonumber \\
&& \nonumber \\
d)\;\;\; && \langle\tilde{\psi}({\bf r},t)\tilde{\psi}({\bf r},t)
\tilde{\psi} ({\bf r},t)\rangle = 0 \nonumber \\
               && \langle\tilde{\psi}^{\dagger}({\bf r},t)
\tilde{\psi}({\bf r},t) \tilde{\psi}({\bf r},t)\rangle = 0 \;\;. \nonumber
\end{eqnarray}
The averages $\langle ... \rangle$ in $a)$, $b)$ and $d)$ are 
nonequilibrium averages, while time-independent equilibrium averages are 
indicated in this paper with the symbol $\langle ... \rangle_0$. 
The prescription $a)$ is the usual decomposition of the field operator 
into a condensate and a noncondensate component and defines the condensate 
wave function $\Phi({\bf r},t)$.
Prescription $b)$ defines the normal, $\tilde{n}({\bf r},t)$, and anomalous, 
$\tilde{m}({\bf r},t)$, noncondensate particle densities. In terms of these 
quantities the interaction term in the Hamiltonian (\ref{gch}) quartic in the 
noncondensate components of $\psi({\bf r},t)$ can be approximated using the 
factorization given by prescription $c)$.   Finally, in prescription 
$d)$ all averages of cubic products of noncondensate operators are set to zero. 
This is expected to be a good approximation for dilute systems. The inclusion of
the triplet correlations $\langle\tilde{\psi}\tilde{\psi}\tilde{\psi}\rangle$ and 
$\langle\tilde{\psi}^{\dagger}\tilde{\psi}\tilde{\psi}\rangle$ in a time-dependent
self-consistent mean-field scheme is discussed in \cite{PB96,RB99}.   
By using the above prescriptions one gets the following equation of 
motion for the condensate wave function
\begin{eqnarray}
i\hbar\frac{\partial}{\partial t}\Phi({\bf r},t) &=& \left(-\frac{\hbar^2
\nabla^2}{2m}+V_{ext}({\bf r})-\mu\right)\Phi({\bf r},t) + 
{\rm g}|\Phi({\bf r},t)|^2
\Phi({\bf r},t) \nonumber \\
&+& 2{\rm g}\Phi({\bf r},t) \tilde{n}({\bf r},t) 
+ {\rm g}\Phi^{\ast}({\bf r},t) \tilde{m}({\bf r},t)\;\;.
\label{gpeq}
\end{eqnarray} 
This equation includes the dynamic coupling between the condensate and 
the noncondensate particles. If we neglect these effects, 
$\tilde{n}=\tilde{m}=0$, equation (\ref{gpeq}) reduces to the usual 
Gross-Pitaevskii (GP) equation.     

We are interested in the small-amplitude oscillations of the condensate, 
which is only slightly displaced from its stationary value
$\Phi_0({\bf r})=\langle\psi({\bf r})\rangle_0$
\begin{equation}
\Phi({\bf r},t)=\Phi_0({\bf r})+\delta\Phi({\bf r},t) \;\;,
\label{flucphi}
\end{equation}
where $\delta\Phi({\bf r},t)$ is a small fluctuation.
In the same way, we consider small fluctuations of the normal and anomalous 
particle densities
\begin{eqnarray}
\tilde{n}({\bf r},t) &=& \tilde{n}^0({\bf r}) 
+ \delta \tilde{n}({\bf r},t) \nonumber \\
\tilde{m}({\bf r},t) &=& \tilde{m}^0({\bf r}) + \delta \tilde{m}({\bf r},t) 
\label{flucden}
\end{eqnarray}
around their  equilibrium values $\tilde{n}^0({\bf r})=\langle\tilde{\psi}
^{\dagger}({\bf r})\tilde{\psi}({\bf r})\rangle_0$ and $\tilde{m}^0({\bf r})=
\langle\tilde{\psi}({\bf r})\tilde{\psi}({\bf r})\rangle_0$.   

The real wave function $\Phi_0({\bf r})$
satisfies the stationary equation \cite{N1}
\begin{equation}
\left(-\frac{\hbar^2\nabla^2}{2m}+V_{ext}({\bf r})-\mu+{\rm g}n_0({\bf r})+
2{\rm g}\tilde{n}^0({\bf r})+{\rm g}\tilde{m}^0({\bf r})\right)\Phi_0({\bf r}) = 0 \;\;,
\label{statgp}
\end{equation}
where
$n_0({\bf r})=|\Phi_0({\bf r})|^2$ is the condensate density. 
The time-dependent equation for $\delta\Phi({\bf r},t)$ is obtained by 
linearizing the equation of motion (\ref{gpeq})
\begin{eqnarray}
i\hbar\frac{\partial}{\partial t}\delta\Phi({\bf r},t) &=& \left( -\frac
{\hbar^2\nabla^2}{2m}+V_{ext}({\bf r})-\mu+2{\rm g} n({\bf r})\right)\delta\Phi
({\bf r},t) \nonumber\\
&+& \left({\rm g}n_0({\bf r})+{\rm g}\tilde{m}^0({\bf r})\right)
\delta\Phi^{\ast}({\bf r},t) 
+ 2{\rm g}\Phi_0({\bf r})\delta
\tilde{n}({\bf r},t) + {\rm g}\Phi_0({\bf r})\delta \tilde{m}({\bf r},t) \;\;,
\label{fluceq}
\end{eqnarray} 
where we have introduced the total equilibrium density
$n({\bf r})=n_0({\bf r})+\tilde{n}^0({\bf r})$.
 
Both the stationary wave function $\Phi_0$ and the fluctuations $\delta\Phi$
depend through Eqs. (\ref{statgp}) and (\ref{fluceq}) on the normal and anomalous 
noncondensate particle densities, for which we need independent equations for their 
equilibrium values and time evolution. To this purpose it is convenient to express
the noncondensate operators $\tilde{\psi}$, $\tilde
{\psi}^{\dagger}$ in terms of quasiparticle operators $\alpha_i$, $\alpha_i^
{\dagger}$ by means of the generalization to inhomogeneous systems of the Bogoliubov 
canonical transformations \cite{F72}
\begin{eqnarray}
\tilde{\psi}({\bf r},t) &=& \sum_i\left(u_i({\bf r})\alpha_i(t) + v_i^{\ast}
({\bf r})\alpha_i^{\dagger}(t)\right) \;\;,
\nonumber \\
\tilde{\psi}^{\dagger}({\bf r},t) &=& \sum_i\left( u_i^{\ast}({\bf r})
\alpha_i^{\dagger}(t) + v_i({\bf r})\alpha_i(t)\right) \;\;.
\label{bogtrans}
\end{eqnarray}
The normalization condition for the functions $u_i({\bf r})$, $v_i({\bf r})$,
which ensures that the quasiparticle operators $\alpha_i$, $\alpha_i^{\dagger}$
satisfy Bose commutation relations, reads
\begin{equation}
\int d{\bf r} \left[u_i^{\ast}({\bf r})u_j({\bf r}) - v_i^{\ast}({\bf r})
v_j({\bf r})\right] = \delta_{ij} \;\;.
\label{norm}
\end{equation}
The time evolution of $\tilde{n}({\bf r},t)$ and $\tilde{m}({\bf r},t)$ can be obtained 
from the Heisenberg equations for the products of quasiparticle operators $\alpha_i^{\dagger}(t)
\alpha_j(t)$ and $\alpha_i(t)\alpha_j(t)$
\begin{eqnarray}
i\hbar\frac{\partial}{\partial t} \langle\alpha_i^{\dagger}(t)\alpha_j(t)\rangle
&=& \langle \Bigl[ 
\alpha_i^{\dagger}
(t)\alpha_j(t),H^{\prime} \Bigr] \rangle \;\;,
\nonumber \\
i\hbar\frac{\partial}{\partial t} \langle\alpha_i(t)\alpha_j(t)\rangle 
&=& \langle \Bigl[ \alpha_i
(t)\alpha_j(t),H^{\prime} \Bigr] \rangle  \;\;.
\label{fgeq}
\end{eqnarray}
In the above equations the commutators are calculated using the mean-field prescriptions
$a)$-$d)$ and the canonical transformation (\ref{bogtrans}). The calculation can be easily 
done by noticing that in the Hamiltonian (\ref{gch}) only the terms quadratic and quartic 
in the noncondensate operators $\tilde\psi$, $\tilde\psi^{\dagger}$ give non vanishing 
contributions, as we set to zero averages of single and cubic products of noncondensate
operators. 

At equilibrium, we take the occupation of quasiparticle levels to be diagonal,
$\langle\alpha_i^\dagger\alpha_j\rangle_0=\delta_{ij}f_i^0$, while anomalous averages of 
quasiparticles are zero, $\langle\alpha_i\alpha_j\rangle_0=0$. With these conditions, the
stationary equations $\langle\Bigl[\alpha_i^{\dagger}(t)\alpha_j(t),H^{\prime} \Bigr]\rangle_0 =
\langle\Bigl[\alpha_i(t)\alpha_j(t),H^{\prime} \Bigr]\rangle_0=0$ yield the following coupled 
equations for the quasiparticle amplitudes $u_i({\bf r})$, $v_i({\bf r})$ 
\begin{eqnarray}
{\cal L}u_i({\bf r})+[{\rm g}n_0({\bf r})+{\rm g}\tilde{m}^0({\bf r})]
v_i({\bf r}) &=& \epsilon_iu_i({\bf r}) \;\;,
\nonumber \\
{\cal L}v_i({\bf r})+[{\rm g}n_0({\bf r})+{\rm g}\tilde{m}^0({\bf r})]
u_i({\bf r}) &=& - \epsilon_i
v_i({\bf r}) \;\;,
\label{bogeqs}
\end{eqnarray}
where we have introduced the hermitian operator
\begin{equation}
{\cal L} = - \frac{\hbar^2\nabla^2}{2m} + V_{ext}({\bf r}) - \mu
+ 2{\rm g}n({\bf r}) \;\;.
\label{Lop}
\end{equation}
The coupled Eqs. (\ref{bogeqs}) correspond to the static Hartree-Fock-Bogoliubov (HFB) equations
as described in Ref. \cite{G96}, and the $\epsilon_i$ are the quasiparticle energies which
fix the quasiparticle occupation numbers at equilibrium $f_i^0=[e^{\epsilon_i/k_BT}-1]^{-1}$.
The equilibrium values of the normal and anomalous particle densities are  
written as
\begin{eqnarray}
\tilde{n}^0({\bf r}) &=& \sum_i \left\{ [|u_i({\bf r})|^2+|v_i({\bf r})|^2]f_i^0
+ |v_i({\bf r})|^2 \right\} \;\;,
\nonumber\\ 
\tilde{m}^0({\bf r}) &=& \sum_i \left\{ 2 u_i({\bf r})v_i^{\ast}({\bf r}) f_i^0 
+ u_i({\bf r}) v_i^{\ast}({\bf r}) \right\} \;\;.
\label{eqdens}
\end{eqnarray}

Out of equilibrium we define the following normal and anomalous quasiparticle
distribution function
\begin{eqnarray}
f_{ij}(t) &=& \langle\alpha_i^{\dagger}(t)\alpha_j(t)\rangle - 
\delta_{ij}f_i^0  \;\;,
\nonumber \\
g_{ij}(t) &=& \langle\alpha_i(t)\alpha_j(t)\rangle \;\;,
\label{qpdens}
\end{eqnarray} 
in terms of which the fluctuations of $\tilde{n}$ and $\tilde{m}$ take the 
form  
\begin{eqnarray}
\delta\tilde{n}({\bf r},t)&=&\sum_{ij} \left\{ [u_i^{\ast}({\bf r})u_j({\bf r})
+ v_i^{\ast}({\bf r})v_j({\bf r})]f_{ij}(t)+u_i({\bf r})v_j({\bf r})g_{ij}(t)
+ u_i^{\ast}({\bf r})v_j^{\ast}({\bf r})g_{ij}^{\ast}(t) \right\}  \;\;,
\nonumber\\
\delta\tilde{m}({\bf r},t)&=&\sum_{ij} \left\{ 2 v_i^{\ast}({\bf r})u_j({\bf r})
f_{ij}(t)+u_i({\bf r})u_j({\bf r})g_{ij}(t)
+ v_i^{\ast}({\bf r})v_j^{\ast}({\bf r})g_{ij}^{\ast}(t) \right\} \;\;.  
\label{neqdens}
\end{eqnarray}
Up to linear terms in the fluctuations $\delta\Phi$, $\delta\tilde{n}$ and 
$\delta\tilde{m}$, the equation of motion for $f_{ij}$ gives the result \cite{N2}
\begin{eqnarray}
i\hbar\frac{\partial}{\partial t}f_{ij}(t) 
&=& (\epsilon_j-\epsilon_i)f_{ij}(t) \,+\, 2{\rm g}(f_i^0-f_j^0) 
\nonumber \\
&\times& \int d{\bf r} \; \Phi_0
\biggl[ \delta\Phi(t) \biggl(u_iu_j^{\ast}+v_iv_j^{\ast}+v_iu_j^{\ast}\biggr)
+ \delta\Phi^{\ast}(t) \biggl(u_iu_j^{\ast}+v_iv_j^{\ast}+u_iv_j^{\ast}\biggr)
\biggr]  
\label{feq} \\
&+& {\rm g}(f_i^0-f_j^0)\int d{\bf r}  \biggl[ 2\delta\tilde{n}(t)
\biggl(u_iu_j^{\ast}+v_iv_j^{\ast}\biggr) +\delta\tilde{m}(t)v_iu_j^\ast
+\delta\tilde{m}^\ast(t)u_iv_j^\ast\biggr] \;\;. \nonumber
\end{eqnarray}
Analogously, for the time evolution of the anomalous quasiparticle distribution 
function $g_{ij}$, one obtains in the linear regime 
\begin{eqnarray}
i\hbar\frac{\partial}{\partial t}g_{ij}(t) 
&=& (\epsilon_j+\epsilon_i)g_{ij}(t) \,+\, 2{\rm g}(1+f_i^0+f_j^0) 
\nonumber \\
&\times& \int d{\bf r} \; \Phi_0
\biggl[ \delta\Phi(t) \biggl(u_i^{\ast}v_j^{\ast}+v_i^{\ast}u_j^{\ast}
+u_i^{\ast}u_j^{\ast}\biggr)
+ \delta\Phi^{\ast}(t) \biggl(u_i^{\ast}v_j^{\ast}+v_i^{\ast}u_j^{\ast}
+v_i^{\ast}v_j^{\ast}\biggr)
\biggr] 
\label{geq} \\
&+& {\rm g}(1+f_i^0+f_j^0)\int d{\bf r}  \biggl[ 2\delta\tilde{n}(t)
\biggl(u_i^{\ast}v_j^{\ast}+v_i^{\ast}u_j^{\ast}\biggr) 
+\delta\tilde{m}(t)u_i^{\ast}u_j^\ast
+\delta\tilde{m}^\ast(t)v_i^{\ast}v_j^\ast\biggr] \;\;.\nonumber
\end{eqnarray}

The first term on the r.h.s. of Eqs. (\ref{feq}), (\ref{geq}) describes
the free evolution of the quasiparticle states, corresponding to quasiparticle 
operators which evolve in time according to $\alpha_i(t)=e^{-i\epsilon_i t/\hbar}
\alpha_i$. The second and the third term describe, respectively, the coupling to the 
oscillations of the condensate and to the fluctuations $\delta\tilde{n}$ and 
$\delta\tilde{m}$ of the normal and anomalous particle density.   
Above the  Bose-Einstein transition temperature, where the system becomes normal and
$\Phi=\tilde{m}=0$, Eq. (\ref{feq}) corresponds to the linearized time-dependent 
Hartree-Fock (TDHF) equation. In the semiclassical limit TDHF is equivalent to the 
collisionless Boltzmann equation for the particle distribution function \cite{RS80}.

In the framework of mean-field theories, coupled time-dependent equations for the 
condensate and noncondensate components of a Bose gas have been discussed by many 
authors. The TDHF scheme is discussed in \cite{MT97}. 
Moreover, coupled equations of motion for the condensate and for correlation functions 
of pairs and triplets of noncondensate particles have been derived in \cite{PB96},
and studied in the linear response regime in \cite{RB99}. Similar kinetic equations
were derived by Kane and Kadanoff \cite{KK65} using the formalism of non-equilibrium 
Green's functions developed in \cite{KB62}. Recently, the approach of Kane and Kadanoff
has been extended to deal with a trapped Bose-condensed gas in \cite{ITG99}.   
 
The stationary Eq. (\ref{statgp}) for $\Phi_0$,
with the normal and anomalous particle densities at equilibrium given by 
(\ref{eqdens}), and Eqs. (\ref{bogeqs}) for the quasiparticles correspond
to the static self-consistent HFB approximation as reviewed by Griffin in 
\cite{G96}. These equations have been solved for harmonically trapped 
gases in \cite{HDB98}. As it is well known \cite{GA59}, in the case of homogeneous
systems, the HFB excitation energies exhibit an unphysical gap at long 
wavelengths, which is fixed by the anomalous particle density: 
$\Delta=2{\rm g}\sqrt{n_0|\tilde{m}^0|}$ (see Ref. \cite{G96}).
 
If one neglects the anomalous particle density, $\tilde{m}^0=0$, the 
HFB equations reduce to the so-called HFB-Popov approximation \cite{P65,G96,SG98},
which is gapless and in the high temperature regime coincides with the Hartree-Fock 
scheme \cite{GSL81}. The HFB-Popov approximation has been applied by several authors
to interacting bosons in harmonic traps, both to calculate the frequencies of the 
collective modes \cite{HZG97} and to study the thermodynamic properties of the system 
\cite{GPS96,GPS97,DGGPS97}. Gapless static mean-field approximations, alternative to 
HFB-Popov, have been put forward and discussed in \cite{PMCB98,RB99}.

Finally, by neglecting both the normal and the anomalous particle density, 
$\tilde{n}^0=\tilde{m}^0=0$, the HFB equations reduce to the Gross-Pitaevskii theory. 
From Eq. (\ref{statgp}) one recovers the stationary GP equation, while Eqs. (\ref{bogeqs}) 
follow from the linearization of the time-dependent GP equation. At very low temperatures,
where effects arising from the depletion of the condensate are negligible, the Gross-Pitaevskii 
theory is well suited to describe dilute gases in traps. For these systems the linearized GP equation 
has been solved by many authors \cite{ERBDC96,EDCB96,DGGPS97} and successfully compared to 
experiments.  

The linearized TDHFB mean-field approximation is a closed set of self-consistent equations
which describe the small oscillations of the 
system around the static HFB solution. The dynamic Eq. (\ref{fluceq}) describes the fluctuation 
$\delta\Phi$ of the condensate around the stationary solution $\Phi_0$, while Eqs. 
(\ref{neqdens})-(\ref{geq}) describe the small oscillations 
$\delta\tilde{n}$, $\delta\tilde{m}$ of the normal and anomalous particle density around
their equilibrium values $\tilde{n}^0$, $\tilde{m}^0$. Since the equations for the time 
evolution of $\delta\Phi$, $\delta\tilde{n}$ and $\delta\tilde{m}$ are derived from the
corresponding exact Heisenberg equations, it can be easily checked that the linearized TDHFB 
approach preserves important conservation laws, such as number of particles and energy conservation.
This is a general feature of time-dependent mean-field approximations \cite{RS80,BR86}. 
Another important feature of linearized TDHFB is that, even though the quasiparticle energies 
obtained from Eqs. (\ref{bogeqs}) exhibit a gap 
at long wavelengths: $\epsilon_p\to\Delta$ as $p\to 0$, the self-consistent solution of 
Eq. (\ref{fluceq}) is gapless. In fact, one can show that 
this self-consistent solution satisfies the Hugenholtz-Pines theorem \cite{HP59}. 

There are some questions one should address before embarking on the difficult task of 
a self-consistent solution of the linearized TDHFB equations. A first problem concerns the 
equilibrium anomalous density $\tilde{m}^0$ [see Eq. (\ref{eqdens})], which is ultraviolet 
divergent. To second order in the interaction and for homogeneous systems the ultraviolet divergence
is canceled by the renormalization of the coupling constant (see e.g. \cite{P72}): 
${\rm g}\to{\rm g}\left(1+{\rm g}\frac{1}{V}\sum_{\bf p}m/p^2\right)$. How to include 
the renormalization of ${\rm g}$ in a self-consistent calculation  and how to extend this 
renormalization procedure to inhomogeneous systems is still
an open problem. Recently, Burnett and coworkers \cite{PMCB98,RB99} have shown that there is no 
need of renormalization of ${\rm g}$ if one uses, instead of a contact potential, an effective 
interaction, the many-body T-matrix, which includes self-consistently the effects arising from the 
anomalous density. Another problem concerns the gap exhibited by the quasiparticle energies in 
Eqs. (\ref{bogeqs}). The self-consistent solution of these equations defines the equilibrium state of
the system: it fixes the noncondensate densities $\tilde{n}^0$ and $\tilde{m}^0$ through 
Eqs. (\ref{eqdens}), 
and the chemical potential $\mu$ and the condensate wavefunction $\Phi_0$ through Eq. (\ref{statgp}).
Even though the small oscillations around the static HFB solutions 
give rise to a gapless spectrum for the fluctuations of the condensate, properties 
such as the phonon velocity and their damping rate will be affected by an incorrect description 
of the system at equilibrium, originating from the unphysical gap $\Delta$ in the quasiparticle 
energies. In particular, if $k_BT\sim\Delta$, one expects a strong influence of the gap on the 
temperature dependence of these properties. In the present work we will not solve the linearized
TDHFB equations self-consistently, instead, we solve them perturbativelly to order ${\rm g}^2$. 
By this method we avoid the problems mentioned above, in particular, the quasiparticle states 
in (\ref{bogeqs}) are properly described by Bogoliubov theory, which is gapless.
 
Another point which deserves some comments is the Kohn mode (dipole mode). As it is well known,
in the presence of harmonic confinement the center of mass degrees of freedom separate from all
other degrees of freedom, giving rise to a collective mode of the system, the Kohn mode, in which 
the equilibrium density profile oscillates rigidly at the trap frequency. The linearized TDHFB 
equations obtained in this section do not describe this mode, as they do not account for the motion
of the center of mass. In fact, these equations correctly describe excitations for which the center 
of mass is at rest, and we will restrict our analysis to this class of excitations. The Kohn mode is 
associated with broken translational symmetry and is often referred 
to as a ``spurious'' mode. For a discussion of spurious states and their appearance in linearized 
time-dependent mean-field theories see Ref. \cite{BR86}.

\section{Spatially homogeneous system}

In this section we will develop, starting from the dynamic equations for the condensate 
and noncondensate components derived in the previous section, a perturbation scheme for 
the elementary excitations in a homogenous system. Explicit results for the temperature
dependence of the chemical potential, damping rate and speed of zero 
sound are given in the limit $a^3n_0\ll 1$, where $n_0=n_0(T)$ is the 
equilibrium value of the condensate density at temperature $T$. At zero temperature, the 
calculation presented here is equivalent to the Beliaev second-order approximation of the 
single-particle Green's function \cite{B58}. In the high temperature regime, $k_BT\gg\mu$,
our approach corresponds to the finite-temperature extension of the Beliaev approximation 
recently employed in Refs. \cite{SG98,FS98} (the former reference gives also a systematic 
account of the various perturbation schemes for a uniform Bose gas). The perturbation 
expansion carried out in the present work holds to order $(a^3n_0)^{1/2}$ for any temperature 
in the condensed phase, except clearly very close to the transition, where the time-dependent 
mean-field equations we used as starting point break down. 

In a spatially homogeneous system the condensate wave function at equilibrium is constant, 
$\Phi_0({\bf r})=\sqrt{n_0}$. The stationary equation (\ref{statgp}) reads then
\begin{equation}
\mu = {\rm g}n_0 + 2{\rm g}\tilde{n}^0 + {\rm g}\tilde{m}^0 \;,
\label{homcp}
\end{equation} 
and represents the equation of state of the system, which fixes the chemical potential 
as a function of the condensate density $n_0$ and the temperature $T$.
By using the above result for the chemical potential, equation (\ref{fluceq}) for the 
fluctuations of the condensate becomes
\begin{eqnarray}
i\hbar\frac{\partial}{\partial t}\delta\Phi({\bf r},t) &=& \left( -\frac
{\hbar^2\nabla^2}{2m}+{\rm g} n_0-{\rm g} \tilde{m}^0 \right)\delta\Phi
({\bf r},t) 
+ \left({\rm g}n_0+{\rm g}\tilde{m}^0\right)
\delta\Phi^{\ast}({\bf r},t) 
\nonumber\\ 
&+& 2{\rm g}\sqrt{n_0}\delta
\tilde{n}({\bf r},t) + {\rm g}\sqrt{n_0}\delta \tilde{m}({\bf r},t) \;\;.
\label{homfc}
\end{eqnarray} 
In the above equation the terms involving the equilibrium anomalous density $\tilde{m}^0$ 
account for the coupling between the fluctuations of the condensate and the static distribution 
of noncondensate atoms, while the terms containing $\delta\tilde{n}$ and $\delta\tilde{m}$
describe the dynamic coupling between the condensate and the fluctuations of the noncondensate 
component.

\subsection{Perturbation scheme}

The perturbation scheme applied to Eqs. (\ref{homcp}), (\ref{homfc}) consists in treating 
the terms which give the static and dynamic coupling to the noncondensate component to {\it second} 
order in ${\rm g}$. It means that the static and fluctuating parts of the normal and 
anomalous density need to be calculated only to {\it first} order in 
{\rm g}. To accomplish this task one must retain in the equations for the quasiparticles 
(\ref{bogeqs}), (\ref{feq}) and (\ref{geq}) only the terms which describe the coupling to the 
condensate and neglect all terms arising from the coupling to the noncondensate component. 

Let us suppose that the condensate oscillates with frequency $\omega$ and wave vector 
${\bf q}/\hbar$
\begin{equation}
\delta\Phi({\bf r},t) = \frac{\delta\Phi_1({\bf q})}{\sqrt{V}} e^{i{\bf q}\cdot{\bf r}/\hbar} 
e^{-i\omega t} 
\;\;\;, \;\;\;\;
\delta\Phi^{\ast}({\bf r},t) = \frac{\delta\Phi_2({\bf q})}{\sqrt{V}} 
e^{i{\bf q}\cdot{\bf r}/\hbar} e^{-i\omega t} \;\;.
\label{homfou}
\end{equation} 
Furthermore, the quasiparticle amplitudes be described by plane-wave functions
\begin{equation}
u_{\bf p}({\bf r})=\frac{u_p}{\sqrt{V}}e^{i{\bf p}\cdot{\bf r}/\hbar} \;\;\;,
v_{\bf p}({\bf r})=\frac{v_p}{\sqrt{V}}e^{i{\bf p}\cdot{\bf r}/\hbar} \;\;.
\label{hompw}
\end{equation}
To first order in ${\rm g}$ the quasiparticle equations (\ref{bogeqs}) become
\begin{eqnarray}
(p^2/2m + {\rm g}n_0)u_p + {\rm g}n_0 v_p &=& \epsilon_p u_p \;\;,
\nonumber \\
(p^2/2m + {\rm g}n_0)v_p + {\rm g}n_0 u_p &=& - \epsilon_p v_p \;\;.
\label{hombe}
\end{eqnarray}
These coupled equations coincide with the well-known Bogoliubov equations for the real 
quasiparticle amplitudes $u_p$, $v_p$, which satisfy the following relations
\begin{eqnarray}
&& u_p^2=1+v_p^2=\frac{(\epsilon_p^2+{\rm g}^2n_0^2)^{1/2}+\epsilon_p}
{2\epsilon_p} \;\;,
\nonumber \\
&& u_pv_p = - \frac{{\rm g}n_0}{2\epsilon_p} \;\;,
\label{homuv}
\end{eqnarray} 
with the quasiparticle energy $\epsilon_p$ given by the Bogoliubov spectrum 
\begin{equation}
\epsilon_p = \left(\left(\epsilon_p^0+{\rm g}n_0\right)^2
-{\rm g}^2n_0^2\right)^{1/2} \;\;,
\label{hombs}
\end{equation}
being $\epsilon_p^0=p^2/2m$ the free-particle energy.
Notice that, by employing the equation of state (\ref{homcp}), the full HFB equations 
(\ref{bogeqs}) would coincide with the matrix equation  
(\ref{hombe}) apart from a term ${\rm g}\tilde{m}_0$. This term would appear with the 
{\it minus} sign in the diagonal term and with the {\it plus} sign in the off-diagonal term, 
and is responsible for the gap in $\epsilon_p$ as $p\to 0$. Since we use the Bogoliubov 
result (\ref{hombs}), we avoid the problem of the gap in the quasiparticle spectrum.  

In the same approximation one must neglect in Eqs. (\ref{feq}) and (\ref{geq}) the terms which
describe the coupling to the fluctuations $\delta\tilde{n}$ and $\delta\tilde{m}$ of the 
noncondensate component. Due to the coupling to the 
condensate, which acts as a time-dependent external drive, the only elements of the matrices
$f_{{\bf p},{\bf p}^{\prime}}$, $g_{{\bf p},{\bf p}^{\prime}}$ and $g^{\ast}_{{\bf p},{\bf p}
^{\prime}}$ which oscillate at the frequency $\omega$ are given by
\begin{eqnarray}
f_{{\bf p},{\bf q}+{\bf p}}(\omega) &=& {\rm g}\frac{\sqrt{n_0}}{\sqrt{V}} 
\frac{f_p^0-f_{|{\bf q}+{\bf p}|}^0}{\hbar\omega+(\epsilon_p-\epsilon_{|{\bf q}+{\bf p}|})
+i0}  
\biggl[\left(\delta\Phi_1-\delta\Phi_2\right)
\left(v_pu_{|{\bf q}+{\bf p}|}-u_pv_{|{\bf q}+{\bf p}|}\right) \biggr.
\nonumber\\
&+& \biggl.\left(\delta\Phi_1+\delta\Phi_2\right)
\left(2u_pu_{|{\bf q}+{\bf p}|}+2v_pv_{|{\bf q}+{\bf p}|}+v_pu_{|{\bf q}+{\bf p}|}
+u_pv_{|{\bf q}+{\bf p}|}\right)\biggr] \;\;,
\nonumber\\
g_{{\bf p},{\bf q}-{\bf p}}(\omega) &=& {\rm g}\frac{\sqrt{n_0}}{\sqrt{V}} 
\frac{1+f_p^0+f_{|{\bf q}-{\bf p}|}^0}{\hbar\omega-(\epsilon_p+\epsilon_{|{\bf q}-{\bf p}|})
+i0}  
\biggl[\left(\delta\Phi_1-\delta\Phi_2\right)
\left(u_pu_{|{\bf q}-{\bf p}|}-v_pv_{|{\bf q}-{\bf p}|}\right) \biggr.
\nonumber\\
&+& \biggl.\left(\delta\Phi_1+\delta\Phi_2\right)
\left(2u_pv_{|{\bf q}-{\bf p}|}+2v_pu_{|{\bf q}-{\bf p}|}+u_pu_{|{\bf q}-{\bf p}|}
+v_pv_{|{\bf q}-{\bf p}|}\right)\biggr] \;\;,
\label{homfg} \\
g^{\ast}_{{\bf p},-{\bf q}-{\bf p}}(\omega) &=& {\rm g}\frac{\sqrt{n_0}}{\sqrt{V}} 
\frac{1+f_p^0+f_{|{\bf q}+{\bf p}|}^0}{\hbar\omega+(\epsilon_p+\epsilon_{|{\bf q}+{\bf p}|})
+i0}  
\biggl[\left(\delta\Phi_1-\delta\Phi_2\right)
\left(u_pu_{|{\bf q}+{\bf p}|}-v_pv_{|{\bf q}+{\bf p}|}\right) \biggr.
\nonumber\\
&-& \biggl.\left(\delta\Phi_1+\delta\Phi_2\right)
\left(2u_pv_{|{\bf q}+{\bf p}|}+2v_pu_{|{\bf q}+{\bf p}|}+u_pu_{|{\bf q}+{\bf p}|}
+v_pv_{|{\bf q}+{\bf p}|}\right)\biggr] \;\;.
\nonumber
\end{eqnarray}
In the above equations the frequency $\omega$ has been chosen with an infinitesimally small 
component on the positive imaginary axis. 

As it is well known (see e.g. Ref. \cite{P72}), to treat consistently to order ${\rm g}^2$ 
the properties of a Bose-condensed gas one must include the proper renormalization of the 
coupling constant ${\rm g}\to{\rm g}\left(1+{\rm g}\frac{1}{V}\sum_{\bf p}m/p^2\right)$.
This renormalization of ${\rm g}$ is crucial because it cancels exactly the large-$p$ 
ultraviolet divergence exhibited by the equilibrium anomalous density $\tilde{m}^0$. 
In fact, by using the renormalized value of ${\rm g}$, the term ${\rm g}n_0+{\rm g}\tilde{m}^0$
present in Eqs. (\ref{homcp}), (\ref{homfc}) becomes 
\begin{equation}
{\rm g}n_0+{\rm g}\tilde{m}^0 \to {\rm g}n_0+{\rm g}^2n_0\frac{1}{V}\sum_{\bf p}
\left( \frac{m}{p^2}-\frac{1+2f_p^0}{2\epsilon_p} \right) \;\;,
\label{homrg}
\end{equation}
and is well behaved at large $p$.

To order ${\rm g}^2$, Eq. (\ref{homfc}) for the fluctuations of the condensate
can be finally written in the form
\begin{eqnarray}
\hbar\omega \left(\delta\Phi_1+\delta\Phi_2\right) &=& 
\frac{q^2}{2m}\left(\delta\Phi_1-\delta\Phi_2\right) 
+ \frac{\Sigma_{11}({\bf q},\omega)-\Sigma_{11}({\bf q},-\omega)}{2}
\left(\delta\Phi_1+\delta\Phi_2\right)
\nonumber\\
&+& \left[\frac{\Sigma_{11}({\bf q},\omega)+\Sigma_{11}({\bf q},-\omega)}{2}
-\Sigma_{12}({\bf q},\omega)\right]\left(\delta\Phi_1-\delta\Phi_2\right) \;\;,
\nonumber\\
\hbar\omega \left(\delta\Phi_1-\delta\Phi_2\right) &=& \left(\frac{q^2}{2m} + 2{\rm g}n_0\right)
\left(\delta\Phi_1+\delta\Phi_2\right) 
+ \frac{\Sigma_{11}({\bf q},\omega)-\Sigma_{11}({\bf q},-\omega)}{2}
\left(\delta\Phi_1-\delta\Phi_2\right)
\label{homg2}\\
&+& \left[\frac{\Sigma_{11}({\bf q},\omega)+\Sigma_{11}({\bf q},-\omega)}{2}
+\Sigma_{12}({\bf q},\omega)\right]\left(\delta\Phi_1+\delta\Phi_2\right) \;\;.
\nonumber
\end{eqnarray} 
In the above equations the self-energies $\Sigma_{11}({\bf q},\omega)$ and 
$\Sigma_{12}({\bf q},\omega)$ are proportional to ${\rm g}^2$. They are obtained from 
Eq. (\ref{homfc})
through the expressions (\ref{neqdens}), which give $\delta\tilde{n}$ and $\delta\tilde{m}$ 
in terms of the 
matrices $f_{{\bf p},{\bf p}^{\prime}}$ and $g_{{\bf p},{\bf p}^{\prime}}$, and through Eqs. 
(\ref{homfg}) and (\ref{homrg}). After some straightforward algebra one finds for the self-energy
$\Sigma_{11}$
\begin{eqnarray}
\Sigma_{11}({\bf q},\omega) &=& {\rm g}^2n_0 \frac{1}{V}\sum_{\bf p}
\left(\frac{m}{p^2}-\frac{1+2f_p^0}{2\epsilon_p}\right)
+{\rm g}^2n_0 \frac{1}{V}\sum_{\bf p}\left(f_p^0-f_k^0\right)
\nonumber\\
&\times& \left( \frac{2B_pA_k+8C_pA_k+4B_pB_k+4C_pC_k}{\hbar\omega+(\epsilon_p-\epsilon_k)+i0}
-\frac{2A_pB_k+8C_pB_k+4A_pA_k+4C_pC_k}{\hbar\omega-(\epsilon_p-\epsilon_k)+i0}\right)
\nonumber\\
&+& {\rm g}^2n_0 \frac{1}{V}\sum_{\bf p}\frac{1+2f_p^0}{\epsilon_p}
+{\rm g}^2n_0 \frac{1}{V}\sum_{\bf p}\left(1+f_p^0+f_k^0\right)
\label{homA}\\
&\times& \left( \frac{2A_pA_k+8C_pA_k+4A_pB_k+4C_pC_k}{\hbar\omega-(\epsilon_p+\epsilon_k)+i0}
-\frac{2B_pB_k+8C_pB_k+4B_pA_k+4C_pC_k}{\hbar\omega+(\epsilon_p+\epsilon_k)+i0}\right) \;\;,
\nonumber
\end{eqnarray}
where we have introduced the notations: ${\bf k}={\bf q}+{\bf p}$, $A_p=u_p^2$, $B_p=v_p^2$ 
and $C_p=u_pv_p$. Analogously for $\Sigma_{12}$ one has
\begin{eqnarray}
\Sigma_{12}({\bf q},\omega) &=& {\rm g}^2n_0 \frac{1}{V}\sum_{\bf p}
\left(\frac{m}{p^2}-\frac{1+2f_p^0}{2\epsilon_p}\right)
+{\rm g}^2n_0 \frac{1}{V}\sum_{\bf p}\left(f_p^0-f_k^0\right)
\nonumber\\
&\times& \left( \frac{4C_pB_k+4C_pA_k+4B_pB_k+6C_pC_k}{\hbar\omega+(\epsilon_p-\epsilon_k)+i0}
-\frac{4C_pA_k+4C_pB_k+4A_pA_k+6C_pC_k}{\hbar\omega-(\epsilon_p-\epsilon_k)+i0}\right)
\nonumber\\
&+& {\rm g}^2n_0 \frac{1}{V}\sum_{\bf p}\left(1+f_p^0+f_k^0\right)
\label{homB}\\
&\times& \left( \frac{4C_pB_k+4C_pA_k+4A_pB_k+6C_pC_k}{\hbar\omega-(\epsilon_p+\epsilon_k)+i0}
-\frac{4C_pB_k+4C_pA_k+4B_pA_k+6C_pC_k}{\hbar\omega+(\epsilon_p+\epsilon_k)+i0}\right) \;\;.
\nonumber
\end{eqnarray}
The above expressions for $\Sigma_{11}$ and $\Sigma_{12}$ coincide with the second-order 
self-energies explicitly calculated at finite temperature by Shi and Griffin using diagrammatic
methods \cite{SG98}. 
At zero temperature they correspond to Beliaev's results \cite{B58}, while in the 
high-temperature regime, $k_BT\gg{\rm g}n_0$, they have been recently discussed by Fedichev 
and Shlyapnikov \cite{FS98}.   

By neglecting in (\ref{homg2}) the terms proportional to the self-energies, one is left with 
the equations for 
the fluctuations of the condensate to first order in ${\rm g}$. These equations coincide with 
the quasiparticle equations (\ref{hombe}). The solution is then given by $\delta\Phi_1({\bf q})
=u_q$ and $\delta\Phi_2({\bf q})=v_q$, with $u_q$ and $v_q$ given by (\ref{homuv}). The 
excitation energy is given by the Bogoliubov spectrum (\ref{hombs}).   
To order ${\rm g}^2$ one writes
\begin{equation}
\hbar\omega = \epsilon_q + \delta E - i\gamma \;\;,
\label{homee}
\end{equation}
where $\delta E$ is the real part of the frequency shift and $\gamma$ is the damping rate.
It is straightforward to obtain the second-order correction to $\hbar\omega$ from Eqs. (\ref{homg2}),
one finds 
\begin{equation}
\delta E - i\gamma = \Sigma_{11}({\bf q},\epsilon_q)u_q^2+2\Sigma_{12}({\bf q},\epsilon_q)u_qv_q
+\Sigma_{11}({\bf q},-\epsilon_q)v_q^2 \;\;,
\label{homde}
\end{equation}
where the self-energies have been calculated for $\hbar\omega=\epsilon_q$.
After some algebra one can cast Eq. (\ref{homde}) in the more convenient form
\begin{eqnarray}
\delta E - i\gamma &=& (u_q+v_q)^2\; {\rm g}^2n_0\frac{1}{V}\sum_{\bf p}\frac{m}{p^2}
- (u_q-v_q)^2\;{\rm g}\tilde{m}^0
\nonumber\\
&+& 4{\rm g}^2 \frac{1}{V}\sum_{\bf p}\left(f_p^0-f_k^0\right)
\left(\frac{A_{p,k}^2}{\epsilon_q+(\epsilon_p-\epsilon_k)
+i0}\right)
\label{homde1}\\
&+& 2{\rm g}^2 \frac{1}{V}\sum_{\bf p}\left(1+f_p^0+f_k^0\right)
\left(\frac{B_{p,k}^2}{\epsilon_q-(\epsilon_p+\epsilon_k)
+i0}-\frac{\tilde{B}_{p,k}^2}{\epsilon_q+(\epsilon_p+\epsilon_k)
+i0}\right)\;,
\nonumber
\end{eqnarray}
where we use ${\bf k}={\bf q}+{\bf p}$ and we have introduced the matrices 
\begin{eqnarray}
A_{p,k} &=& \frac{\sqrt{n_0}}{2} \left[ (u_q+v_q) (2u_pu_k+2v_pv_k+v_pu_k+u_pv_k)
+ (u_q-v_q) (v_pu_k-u_pv_k) \right] \;\;,
\nonumber\\
B_{p,k} &=& \frac{\sqrt{n_0}}{2} \left[ (u_q+v_q) (2u_pv_k+2v_pu_k+u_pu_k+v_pv_k)
+ (u_q-v_q) (u_pu_k-v_pv_k) \right] \;\;,
\label{homAB}\\
\tilde{B}_{p,k} &=& \frac{\sqrt{n_0}}{2} \left[ (u_q+v_q) (2u_pv_k+2v_pu_k+u_pu_k+v_pv_k)
- (u_q-v_q) (u_pu_k-v_pv_k) \right] \;\;.
\nonumber
\end{eqnarray}

Result (\ref{homde1}) gives the correction to the Bogoliubov elementary
excitation energy $\epsilon_q$. The first and second term on the r.h.s. of (\ref{homde1}) arise, 
respectively, 
from the renormalization of ${\rm g}$ and from the equilibrium anomalous density $\tilde{m}^0$.
The last two terms arise, instead, from the dynamic coupling between the condensate and the 
noncondensate component. The real part of the r.h.s. of (\ref{homde1}) gives 
the frequency shift $\delta E$, while the imaginary part gives the damping rate $\gamma$. 
Notice that, concerning the damping rate, result (\ref{homde1}) coincides with the calculation
carried out 
within the Popov approximation [Eq. (39) of \cite{G98}], where the condition $\tilde{m}^0=0$
was assumed. However, as discussed in \cite{G98}, the frequency shift $\delta E$ is not given
correctly to order ${\rm g}^2$ by the Popov approximation. In fact, the static anomalous density
$\tilde{m}^0$ and the renormalized coupling constant contribute to the real part of Eq. (\ref{homde1}).

\subsection{Equation of state}

Let us first discuss the equation of state (\ref{homcp}). By calculating $\tilde{n}^0$ and 
$\tilde{m}^0$ using the equilibrium expressions (\ref{eqdens}) with the first-order quasiparticle
amplitudes and energies given by (\ref{homuv}) and (\ref{hombs}), one finds to order ${\rm g}^2$
\begin{equation}
\mu= {\rm g}n_0 + 2{\rm g}n_T^0 + {\rm g}n_0 (a^3n_0)^{1/2} H(\tau) \;\;.
\label{homdcp}
\end{equation}
In the above equation $n_T^0=\zeta(3/2)\lambda_T^{-3}\simeq 2.612\lambda_T^{-3}$ is the noncondensate 
density of an ideal gas, which is fixed by the thermal wavelength $\lambda_T=\hbar\sqrt{2\pi/mk_BT}$.  
Moreover, $H(\tau)$ is a dimensionless function
of the reduced temperature $\tau=k_BT/{\rm g}n_0$ given by
\begin{equation}
H(\tau) = \frac{40}{3\sqrt{\pi}} + \frac{\sqrt{32}}{\sqrt{\pi}}\tau\int_0^{\infty} dx
\frac{1}{e^x-1}\left(\sqrt{u-1}\frac{2u-1}{u}-2\sqrt{\tau x}\right) \;\;,
\label{homH}
\end{equation}
where we have introduced the quantity $u=\sqrt{1+\tau^2x^2}$.
Result (\ref{homdcp}) gives the chemical potential as a function of the equilibrium condensate 
density $n_0$ and the temperature $T$ to second order in ${\rm g}$. It coincides with the  
result of Shi and Griffin [Eq. (7.9) of \cite{SG98}].    
Notice that the sum of the first two terms on the r.h.s. of (\ref{homdcp}) 
corresponds to the chemical potential calculated to first order in ${\rm g}$.
In Fig. 1 the dimensionless function $H(\tau)$ is plotted as a function of the reduced temperature
$\tau$.

At low temperatures, $\tau\ll 1$, the function $H(\tau)$ can be expanded as   
\begin{equation}
H(\tau) \simeq \frac{40}{3\sqrt{\pi}} - \sqrt{32}\zeta(3/2)\tau^{3/2} + \frac{2\pi^{3/2}}{3}\tau^2
+ \frac{\pi^{7/2}}{10}\tau^4 \;\;.
\label{homHLT}
\end{equation}  
In the same regime of temperatures, the condensate density is given in terms of the total 
density $n=n_0+\tilde{n}^0$ by the following relation
\begin{equation}
n_0 \simeq n\left[1-(a^3n_0)^{1/2}\left(\frac{8}{3\sqrt{\pi}}+\frac{2\pi^{3/2}}{3}\tau^2
-\frac{\pi^{7/2}}{30}\tau^4 \right)\right] \;\;,
\label{homcLT}
\end{equation}
valid to order ${\rm g}^2$. In the above expression the first term in brackets corresponds to the 
quantum depletion of the condensate, while the other two terms account for the thermal depletion 
caused by phonon-type excitations. By using Eqs. (\ref{homHLT}) and (\ref{homcLT}), 
one gets the following result for the low-temperature behavior of the chemical potential in 
terms of the density $n$
\begin{eqnarray}
\mu &\simeq& {\rm g}n\left[1+(a^3n_0)^{1/2}\left(\frac{32}{3\sqrt{\pi}}+\frac{2\pi^{7/2}}{15}\tau^4
\right)\right]
\nonumber\\
&\simeq& \mu(T=0) + \frac{\pi^2}{60}\frac{(k_BT)^4}{n_0\hbar^3c_B^3} \;\;,
\label{homcpLT}
\end{eqnarray} 
where $\mu(T=0)={\rm g}n[1+32(a^3n_0)^{1/2}/3\sqrt{\pi}]$ is the value of the chemical potential 
at zero  temperature \cite{LP80}, and $c_B=\sqrt{{\rm g}n_0/m}$ is the Bogoliubov velocity of sound.
The $T^4$ term in (\ref{homcpLT}) coincides with the result obtained from the thermodynamic relation
$\mu=(\partial F/\partial N)_{V,T}$, where $F$ is the low-temperature expansion of the free energy 
of a Bose gas \cite{LP80,G98}.  

At high temperatures, $\tau\gg 1$, the function $H(\tau)$ yields the asymptotic result $H(\tau)
\to - 12\sqrt{\pi}\tau$, and for the chemical potential in this regime of temperatures one gets
\begin{equation}
\mu \simeq {\rm g}n_0 + 2{\rm g}n_T^0 - 12\sqrt{\pi}(a^3n_0)^{1/2}k_BT \;\;,
\label{homcpHT}
\end{equation}
which coincides with the result obtained by Popov \cite{P65,SG98,FS98}.

\subsection{Zero sound: damping rate and velocity of propagation}

Zero sound, or Bogoliubov sound, is a collective oscillation of the system in the collisionless
regime, for which the restoring force acting on a given particle comes from the mean-field created
by the other particles. At zero temperature, zero sound coincides with ordinary sound and the 
velocity of propagation $c$ is fixed by the compressibility of the system 
$mc^2=(\partial P/\partial n)_{T=0}$.
At finite temperature one can have both zero sound and hydrodynamic
modes, depending on the temperature and the wavelength of the excitation. In this case the 
velocity of zero sound can not be fixed only by thermodynamic relations. For an exhaustive 
discussion of collisionless and hydrodynamic collective modes we refer to the books \cite{NP90,G93}.

To first order in ${\rm g}$, the excitation energy of the zero-sound mode is given by the 
long-wavelength limit of the Bogoliubov dispersion relation (\ref{hombs}), which corresponds 
to phonons $\epsilon_q=c_B q$ propagating with the velocity
$c_B=\sqrt{{\rm g}n_0/m}$ fixed by the condensate density.
Starting from Eq. (\ref{homde1}), we will explicitly calculate the damping rate and the temperature
dependence of the speed of zero sound to order ${\rm g}^2$. 
The damping of phonons in homogeneous systems has been recently investigated by Liu \cite{L97}, 
using functional-integration methods, and by Pitaevskii and Stringari \cite{PS97} by means of 
perturbation theory. The temperature dependence of the speed of zero sound has been investigated
by Payne and Griffin \cite{PG85} within the framework of the dielectric formalism, and by Shi and 
Griffin \cite{SG98} and Fedichev and Shlyapnikov \cite{FS98} using diagrammatic methods.  
  
\subsubsection{Damping of zero sound}

The calculation of the damping of phonons from Eq. (\ref{homde1}) has been already carried out 
in \cite{G98}. Here we will only review the main results.

In the quantum regime, $\epsilon_q\gg k_BT$, the damping is governed by the Beliaev mechanism, in 
which a phonon decays into a pair of excitations. This mechanism is described in Eq. (\ref{homde1}) 
by the imaginary part of the term containing the matrices $B$ and 
$\tilde{B}$. The damping rate is given by
\begin{equation}
\frac{\gamma}{c_B q}=\frac{3q^4}{640\pi\hbar^3mn_0c_B} \;\;.
\label{homBd}
\end{equation}
This result has been first obtained by Beliaev using diagrammatic techniques \cite{B58}.

In the thermal regime, $\epsilon_q\ll k_BT$, the phonon decay is dominated by a different damping 
process, in which a phonon with energy $\epsilon_q$ is absorbed by a thermal excitation with energy
$\epsilon_p$ and is turned into another thermal excitation with energy $\epsilon_{|{\bf q}+{\bf p}|}$.
This mechanism is known as Landau damping and is described in Eq. (\ref{homde1}) by the imaginary part
of the term involving the matrix $A$. The result is given by (see Refs. \cite{PS97,G98})
\begin{equation}
\frac{\gamma}{c_B q} = (a^3n_0)^{1/2} F(\tau) \;\;,
\label{homLd}
\end{equation}
In the above equation $\tau=k_BT/{\rm g}n_0$ is the reduced temperature and 
$F(\tau)$ is the dimensionless function
\begin{equation}
F(\tau)=4\sqrt{\pi}\int_0^\infty dx \left(e^{x/2}-e^{-x/2}\right)^{-2} 
\left(1-\frac{1}{2u}-\frac{1}{2u^2}\right)^2 \;\;,
\label{homF}
\end{equation}
where $u$ is defined as in (\ref{homH}).

For temperatures $\tau\ll 1$ the function $F$ takes the asymptotic limit 
$F\rightarrow 3\pi^{9/2}\tau^4/5$ and one recovers the known 
result for the phonon damping \cite{AK63,HM65,LP81,PS97} 
\begin{equation}
\frac{\gamma}{c_B q}=\frac{3\pi^3}{40}\frac{(k_BT)^4}{mn_0\hbar^3c_B^5} \;\;.
\label{homKd}
\end{equation}

In the opposite regime of temperatures, $\tau\gg 1$, one finds the asymptotic 
limit $F\rightarrow 3\pi^{3/2}\tau/4$, and the damping coefficient is given by 
\begin{equation}
\frac{\gamma}{c_B q}=\frac{3\pi}{8}\frac{k_BTa}{\hbar c_B} \;\;.
\label{homSd}
\end{equation}
The damping of phonons in this regime of temperatures has been first investigated 
by Sz\'{e}pfalusy and Kondor \cite{SK74}. 

In Fig. 2 the dimensionless function $F(\tau)$ is plotted as a function of
$\tau$ together with its asymptotic behaviour both at small and large  
$\tau$'s. One can see that $F$ departs rather soon from the low-temperature 
$\tau^4$ behaviour, while it approaches the high-temperature  
linear law very slowly.

\subsubsection{Velocity of zero sound}

Differently from the calculation of the damping rates, all terms on the r.h.s. of (\ref{homde1}) 
contribute to the frequency shift $\delta E$. The first two terms, which involve $\tilde{m}_0$ 
and the renormalization of the coupling constant, are referred to as static terms. 
Concerning the other two terms,
for excitations with $\epsilon_q\ll{\rm g}n_0$, the relevant region in the calculation 
of the real part of the term which containes the matrices $B$ and $\tilde{B}$ corresponds to 
energies $\epsilon_p\gg\epsilon_q$. This term is referred to as non-resonant term.
On the contrary, the resonance region gives an important contribution to the real part of the term 
involving the matrix $A$. This term is referred to as resonant term.  

Let us start by analysing the non-resonant term. The contribution of this term to the energy 
shift $\delta E$ can be written as
\begin{eqnarray}
\delta E_{NR} &=& -\;\frac{\epsilon_q}{\epsilon_q^0}\;{\rm g}^2n_0 \frac{1}{V}\sum_{\bf p}
\left[ f_p^0\,\frac{\epsilon_p-\epsilon_k}{2\epsilon_p\epsilon_k} \;+\;
\frac{1+2f_p^0}{2\epsilon_p} \;-\;(1+2f_p^0)\,
\frac{(\epsilon_k^0-\epsilon_p^0)^2}{4\epsilon_p\epsilon_k(\epsilon_p+\epsilon_k)}\right]
\nonumber\\
&+& \epsilon_q\;{\rm g}^2n_0\frac{1}{V}\sum_{\bf p}(1+2f_p^0)\frac{\epsilon_p+\epsilon_k}
{\epsilon_q^2-(\epsilon_p+\epsilon_k)^2}
\left[\frac{\epsilon_q^0}{\epsilon_q^2}(2u_pv_k+2v_pu_k+u_pu_k+v_pv_k)^2\right.
\nonumber\\
&+& \left. 2\frac{u_pu_k-v_pv_k}{\epsilon_p+\epsilon_k}(2u_pv_k+2v_pu_k+u_pu_k+v_pv_k)
+ \frac{\epsilon_q^2}{\epsilon_q^0}\frac{(u_pu_k-v_pv_k)^2}{(\epsilon_p+\epsilon_k)^2}
\right] \;\;,
\label{homENR}
\end{eqnarray}
where ${\bf k}= {\bf q}+{\bf p}$. The above result is valid for any excitation 
energy $\epsilon_q$, and is not limited to the long-wavelength regime $\epsilon_q\ll{\rm g}n_0$.
In the phonon regime, result (\ref{homENR}) can be simplified and one gets
\begin{eqnarray}
\frac{\delta E_{NR}}{c_B q} &=& - {\rm g}\frac{1}{V}\sum_{\bf p} (1+2f_p^0)
\left[\frac{(\epsilon_p^0)^2}{4\epsilon_p^3} - \frac{{\rm g}n_0\epsilon_p^0}{6\epsilon_p^3}\right]
- \frac{1}{\epsilon_q^0} {\rm g}^2n_0\frac{1}{V}\sum_{\bf p}\frac{1+2f_p^0}{2\epsilon_p}
\nonumber\\
&-& \frac{1}{\epsilon_q^0} {\rm g}^2n_0\frac{1}{V}\sum_{\bf p} f_p^0 \frac{\epsilon_p-\epsilon_k}
{2\epsilon_p\epsilon_k} \;\;.
\label{homENR1}
\end{eqnarray}

The contribution to $\delta E$ from the resonant term can be calculated in the same way 
and one gets the general result
\begin{eqnarray}
\delta E_R &=& \frac{\epsilon_q}{\epsilon_q^0} \;{\rm g}^2n_0 \frac{1}{V}\sum_{\bf p}\left[ f_p^0\,
\frac{\epsilon_p-\epsilon_k}{2\epsilon_p\epsilon_k} \;-\; 
\frac{f_p^0-f_k^0}{\epsilon_p-\epsilon_k}\,
\frac{(\epsilon_k^0-\epsilon_p^0)^2}{4\epsilon_p\epsilon_k}\right]
+\epsilon_q\;{\rm g}^2n_0 \frac{1}{V}\sum_{\bf p}\frac{f_p^0-f_k^0}{\epsilon_q+\epsilon_p-\epsilon_k}
\nonumber\\
&\times& \left[\frac{\epsilon_q^0}{\epsilon_q^2}(2u_pu_k+2v_pv_k+v_pu_k+u_pv_k)^2
-2\frac{v_pu_k-u_pv_k}{\epsilon_p-\epsilon_k}(2u_pu_k+2v_pv_k+v_pu_k+u_pv_k)\right.
\nonumber\\
&+& \left. \frac{\epsilon_q^2}{\epsilon_q^0}\frac{(v_pu_k-u_pv_k)^2}{(\epsilon_p-\epsilon_k)^2}
\right] \;\;.
\label{homER}
\end{eqnarray}
In the limit $\epsilon_q\ll{\rm g}n_0$ the above expression reduces to 
\begin{eqnarray}
\frac{\delta E_{R}}{c_B q} &=& 
\frac{\rm g}{2V}\sum_{\bf p}\frac{\partial f_p^0}{\partial\epsilon_p}
\left[ \left(\frac{\epsilon_p^0}{\epsilon_p}+\frac{\partial\epsilon_p^0}{\partial\epsilon_p}
\right)^2\left(1-\frac{c_B}{c_B-\cos\theta \partial\epsilon_p/\partial p}\right)-
\frac{2}{3}\frac{{\rm g}n_0\epsilon_p^0}{\epsilon_p^2}\right]
\nonumber\\
&+& \frac{1}{\epsilon_q^0} {\rm g}^2n_0\frac{1}{V}\sum_{\bf p} f_p^0 \frac{\epsilon_p-\epsilon_k}
{2\epsilon_p\epsilon_k} \;\;,
\label{homER1}
\end{eqnarray}
where $\theta$ is the angle the momentum ${\bf p}$ forms with the direction of ${\bf q}$.

Notice that the last terms on the r.h.s. of (\ref{homENR1}) and (\ref{homER1}) are equal and opposite 
in sign, thus they cancel in the sum $\delta E_R + \delta E_{NR}$. 
Moreover, the second term on the r.h.s. of (\ref{homENR1}), which is ultraviolet divergent, 
is canceled by a corresponding term arising from the contribution to 
$\delta E$ of the static terms. In conclusion, the final result for the energy shift $\delta E$ 
in the phonon regime is well behaved and proportional to $\epsilon_q$. One finds 
\begin{eqnarray}
\frac{\delta E}{c_B q}&=& \frac{\rm g}{2V}\sum_{\bf p}\frac{\partial f_p^0}{\partial\epsilon_p}
\left[ \left(\frac{\epsilon_p^0}{\epsilon_p}+\frac{\partial\epsilon_p^0}{\partial\epsilon_p}
\right)^2\left(1-\frac{c_B}{c_B-\cos\theta \partial\epsilon_p/\partial p}\right)-
\frac{2}{3}\frac{{\rm g}n_0\epsilon_p^0}{\epsilon_p^2}\right]
\nonumber\\
&+& \frac{\rm g}{2V}\sum_{\bf p}\left[ \left(\frac{m}{p^2}-\frac{1}{2\epsilon_p}
+\frac{4}{3}\frac{{\rm g}n_0\epsilon_p^0}{\epsilon_p^3}\right)-\frac{f_p^0}{\epsilon_p}
\left(1-\frac{8}{3}\frac{{\rm g}n_0\epsilon_p^0}{\epsilon_p^2}\right)\right] \;\;.
\label{homEZS}
\end{eqnarray}

The r.h.s. of the above equation gives the correction to the Bogoliubov velocity of zero 
sound. By rearranging the integrals over momenta, one gets the relevant result
\begin{equation}
c=c_B\left[ 1+(a^3n_0)^{1/2} G(\tau) \right] \;\;.
\label{homEZS1}
\end{equation}
Here $G(\tau)$ is the following dimensionless function of 
the reduced temperature $\tau=k_BT/{\rm g}n_0$
\begin{eqnarray}
G(\tau) &=& \frac{28}{3\sqrt{\pi}} \;+\; \frac{\sqrt{32}}{\sqrt{\pi}}\tau
\int_0^\infty dx \frac{1}{e^x-1} \frac{\sqrt{u-1}}{u}\frac{5-3u}{6(u+1)}
\;+\; \frac{\sqrt{32}}{\sqrt{\pi}}\int_0^\infty dx \frac{1}{(e^{x/2}-e^{-x/2})^2}
\nonumber\\
&\times& \frac{u-1}{3u\sqrt{u+1}} 
\left[1-\frac{3}{2}\frac{(2u+1)^2(u-1)}{u^2}
\left(1+\frac{\sqrt{u+1}}{2\sqrt{2}u}\log\left|\frac{\sqrt{2}u-\sqrt{u+1}}{\sqrt{2}u+\sqrt{u+1}}
\right|\right)\right] \;\;,
\label{homG}
\end{eqnarray}
where $u$ is defined as in (\ref{homH}).
 
It is interesting to study Eq. (\ref{homEZS1}) in particular regimes of temperature.  
At zero temperature, the function $G$ takes the value: $G(\tau=0) = 28/(3\sqrt{\pi})$, 
and $n_0$ is related to the total density $n$ by the expression: 
$n_0 = n[1-8/(3\sqrt{\pi})(a^3n_0)^{1/2}]$, which accounts for the quantum depletion.
The result for the sound velocity is:
\begin{equation}
c(T=0)=\sqrt{\frac{{\rm g}n}{m}}\left[ 1+\frac{8}{\sqrt{\pi}}(a^3n_0)^{1/2} \right] \;\;.
\label{cT0}
\end{equation}
The above result, which was first found by Beliaev \cite{B58}, coincides with the one obtained 
from the thermodynamic relation $c(T=0)=[n(\partial\mu(T=0)/\partial n)/m]^{1/2}$, where 
$\mu(T=0)$ is given in (\ref{homcpLT}).  

At low temperatures, $\tau\ll 1$, one finds the following expansion of the $G$ function
\begin{equation}
G(\tau)\simeq\frac{28}{3\sqrt{\pi}} + \frac{\pi^{3/2}}{3}\tau^2 + \frac{3\pi^{7/2}}{5}
\tau^4\log(1/\tau^2) \;\;.
\label{GLT}
\end{equation}
In this regime of temperatures the condensate density is given by the expression (\ref{homcLT})
in terms of the density $n$, and the velocity of zero sound turns out to be 
\begin{equation}
c=c(T=0) + \frac{3\pi^2}{40} \frac{(k_BT)^4}{mn_0\hbar^3c_B^4} \log[m^2c_B^4/(k_BT)^2] \;\;.
\label{cLT}
\end{equation}
This result was first obtained by Andreev and Khalatnikov \cite{AK63} using kinetic equations,
and later by Ma {\it et al.} \cite{MGW71} within the framework of the dielectric formalism.

Finally, in the high temperature regime $\tau\gg 1$, the function $G$ is linear in $\tau$: 
$G(\tau)\to G(\infty) \tau$, with the numerical coefficent $G(\infty)$ given by the 
following expression
\begin{equation}
G(\infty)= \frac{\sqrt{\pi}}{3}(9\sqrt{2}-28) + \frac{1}{\sqrt{\pi}} \int_0^1 dx\, 
\frac{x^3+3x^2-4}{\sqrt{1-x^2}(1+x)} \log\left(\frac{\sqrt{2}-\sqrt{x(1+x)}}{\sqrt{2}+\sqrt{x(1+x)}}
\right) \simeq - 7.4 \;\;.
\label{Ginf}
\end{equation}
For the speed of zero sound in this regime of temperatures one gets the result 
\cite{SG98,FS98}
\begin{equation}
c=c_B+\frac{G(\infty)}{2\sqrt{\pi}} \frac{k_BTa}{\hbar} \;\;,
\label{cHT}
\end{equation}
where the numerical coefficient $G(\infty)/(2\sqrt{\pi})\simeq -2.1$ agrees with the finding 
of \cite{FS98}, while is about a factor 6 larger than the one calculated in \cite{SG98}. 

The proper description of the cross-over between the low and high-temperature regime is provided by 
Eq. (\ref{homEZS1}). The dimensionless function $G(\tau)$ is plotted
in Fig. 3. In the experiments on trapped gases the gas parameter in the center of the trap is 
typically $a^3n_0\sim$ 10$^{-5}$-10$^{-4}$. For temperatures of the order of the chemical potential, 
which means $\tau\sim 1$, the correction to the Bogoliubov speed of sound amounts
to about 2-5\%.

\section{Spatially inhomogeneous system}

In this section we generalize the perturbation scheme developed for a homogeneous Bose-condensed
gas to the case of inhomogenous systems trapped by a harmonic confining potential
\begin{equation}
V_{ext}({\bf r}) = \frac{m}{2}\left(\omega_x^2x^2 + \omega_y^2y^2 + \omega_z^2z^2\right) \;\;.
\label{harp}
\end{equation} 
The relevant length scale associated to the external potential (\ref{harp}) is the harmonic 
oscillator length defined as
\begin{equation}
a_{ho}=\left(\frac{\hbar}{m\omega_{ho}}\right)^{1/2} \;\;,
\label{aho}
\end{equation}
where $\omega_{ho}=(\omega_x\omega_y\omega_z)^{1/3}$ is the geometric average of the oscillator
frequencies. The length scale $a_{ho}$ gives the average width of the Gaussian which describes 
the ground state of non-interacting particles in the harmonic potential (\ref{harp}).
The shape of the potential $V_{ext}$ fixes the symmetry of the problem. So far all experiments 
on trapped Bose gases have been realized using axially symmetric traps. In this case there are 
only two distinct oscillator frequencies: $\omega_{\perp}=\omega_x=\omega_y$ and $\omega_z$.  
The ratio between the axial and radial frequencies, $\lambda=\omega_z/\omega_{\perp}$, fixes the 
asymmetry of the trap. For $\lambda<1$ the trap is cigar shaped, whereas for $\lambda>1$ is disk 
shaped and $\lambda=1$ refers to spherically symmetric traps.

In our analysis we only consider systems with repulsive interactions ($a>0$) in the thermodynamic 
limit. As extensively discussed in \cite{DGPS99}, for harmonically trapped Bose systems the 
thermodynamic limit  is obtained by letting the total number of trapped particles $N\to\infty$ and 
$\omega_{ho}\to 0$, while keeping the product $N\omega_{ho}^3$ constant. 
With this definition the Bose-Einstein transition temperature $k_BT_c^0=\hbar\omega_{ho}(N/\zeta(3))
^{1/3}$ is well defined in the thermodynamic limit. 

In the thermodynamic limit, the condition $N_0(T)a/a_{ho}\gg 1$, which ensures the validity of the 
Thomas-Fermi (TF) approximation for the condensate with occupation number $N_0$, is always guaranteed 
below the transition temperature. In the TF approximation one neglects the quantum-pressure term 
proportional to $\nabla^2\Phi_0({\bf r})$ in the stationary equation (\ref{statgp}), and the 
equilibrium profile of the condensate density is fixed by the following equation 
\begin{equation}
{\rm g}n_0({\bf r})=\mu-V_{ext}({\bf r})-2{\rm g}\tilde{n}^0({\bf r})-{\rm g}\tilde{m}^0({\bf r}) \;\;,
\label{eqcd}
\end{equation}
in the central region of the trap where the r.h.s. of the above equation is positive, whereas 
outside this region one has $n_0({\bf r)}=0$.
The chemical potential in Eq. (\ref{eqcd}) is fixed by the normalization condition $\int d{\bf r}\;
n_0({\bf r})=N_0(T)$, with $N_0(T)$ the equilibrium condensate occupation number at temperature $T$.
To lowest order in the interaction, the profile of the condensate density has the form of the inverted 
parabola
\begin{equation}
n_{TF}({\bf r})={\rm g}^{-1} \left[ \mu_{TF}(N_0)-V_{ext}({\bf r}) \right]  \;\;,
\label{nTF}
\end{equation} 
where
\begin{equation}
\mu_{TF}(N_0)=\frac{\hbar\omega_{ho}}{2}\left(\frac{15 N_0 a}{a_{ho}}\right)^{2/5}
\label{muTF}
\end{equation}
is the temperature dependent TF chemical potential.
Moreover, in the thermodynamic limit, one can show \cite{GPS97,DGPS99} that the equilibrium 
properties of the system can be expressed in terms of two parameters: the reduced temperature 
$t=T/T_c^0$ and the interaction parameter $\eta$ defined as the ratio
\begin{equation}
\eta=\frac{\mu_{TF}(N)}{k_BT_c^0} \;\;,
\label{eta}
\end{equation}    
between the TF chemical potential at $T=0$ and the transition temperature.

The time-dependent equation (\ref{fluceq}) for the fluctuations of the condensate 
only needs to be solved in the region where $n_0({\bf r})\neq 0$, according to Eq. (\ref{eqcd}). 
One finds
\begin{eqnarray}
i\hbar\frac{\partial}{\partial t}\delta\Phi({\bf r},t) &=& \left( -\frac
{\hbar^2\nabla^2}{2m}+{\rm g} n_0({\bf r})-{\rm g} \tilde{m}^0({\bf r}) \right)\delta\Phi
({\bf r},t) 
+ \left({\rm g}n_0({\bf r})+{\rm g}\tilde{m}^0({\bf r})\right)
\delta\Phi^{\ast}({\bf r},t) 
\nonumber\\
&+& 2{\rm g}\Phi_0({\bf r})\delta
\tilde{n}({\bf r},t) + {\rm g}\Phi_0({\bf r})\delta \tilde{m}({\bf r},t) \;\;.
\label{eqcf}
\end{eqnarray}
For a trapped system in the thermodynamic limit the above equations ({\ref{eqcd}) and (\ref{eqcf}) 
replace respectively Eqs. (\ref{homcp}) and (\ref{homfc}), holding for a homogeneous system.

We are interested in the lowest-lying collective modes of the system with excitation energy
$\hbar\omega\ll\mu$. To lowest order in the interaction these modes are the solution of the 
following coupled equations
\begin{eqnarray}
i\hbar\frac{\partial}{\partial t}\left[\delta\Phi({\bf r},t)+\delta\Phi^{\ast}({\bf r},t)\right]
&=& -\frac{\hbar^2\nabla^2}{2m} \left[\delta\Phi({\bf r},t)-\delta\Phi^{\ast}({\bf r},t)\right]
\;\;,
\nonumber\\
i\hbar\frac{\partial}{\partial t}\left[\delta\Phi({\bf r},t)-\delta\Phi^{\ast}({\bf r},t)\right]
&=& 2{\rm g}n_{TF}({\bf r})\left[\delta\Phi({\bf r},t)+\delta\Phi^{\ast}({\bf r},t)\right]
\;\;.
\label{hydeq}
\end{eqnarray}
These equations are obtained from (\ref{eqcf}) by neglecting all coupling terms to noncondensate 
particles and neglecting also the term proportional to $\nabla^2(\delta\Phi+\delta\Phi^{\ast})$,
which is of higher order for the low-lying modes we are considering \cite{WG96}.  

The oscillating solution defined by $\delta\Phi({\bf r},t)=\delta\Phi_1^0({\bf r})e^{-i\omega t}$,
$\delta\Phi^{\ast}({\bf r},t)=\delta\Phi_2^0({\bf r})e^{-i\omega t}$, with the Fourier components
fixed by the relations
\begin{eqnarray}
\left(\delta\Phi_1^0({\bf r})+\delta\Phi_2^0({\bf r})\right) &=& \sqrt{\frac{\hbar\omega}
{2{\rm g}n_{TF}({\bf r})}}\;\chi_0({\bf r}) \;\;,
\nonumber\\
\left(\delta\Phi_1^0({\bf r})-\delta\Phi_2^0({\bf r})\right) &=& \sqrt{\frac{2{\rm g}n_{TF}({\bf r})}
{\hbar\omega}}\;\chi_0({\bf r}) \;\;,
\label{hydcom}
\end{eqnarray}
reduces the coupled equations (\ref{hydeq}) to the following equation for the function 
$\chi_0({\bf r})$ \cite{WG96}
\begin{equation}
m\omega^2\chi_0({\bf r}) + \nabla\Bigl[ {\rm g}n_{TF}({\bf r})\nabla\chi_0({\bf r})\Bigr] = 0
\;\;.
\label{hydeq1}
\end{equation}
The normalization condition $\int d{\bf r}\left(|\delta\Phi_1^0|^2-|\delta\Phi_2^0|^2\right)=1$
satisfied by the Fourier components $\delta\Phi_1^0$ and $\delta\Phi_2^0$, implies the normalization 
condition $\int d{\bf r}\;\chi_0^{\ast}\chi_0^{\,} =1$ on the function $\chi_0({\bf r})$. 

Equation (\ref{hydeq1}) was first derived at $T=0$ by Stringari \cite{S96} using the hydrodynamic 
theory of superfluids, and it has then been studied by many authors \cite{WG96,FCSG97}.
For spherically symmetric traps the excitation energies $\hbar\omega\equiv\epsilon_{TF}$ obey 
the dispersion law \cite{S96}
\begin{equation}
\epsilon_{TF}(n_r,l) = \hbar\omega_{ho} \left( 2n_r^2+2n_rl+3n_r+l \right)^{1/2} \;\;,
\label{symdl}
\end{equation}
where $n_r$ and $l$ are respectively the radial and the angular momentum quantum numbers.
In the case of axially symmetric traps, analytic results for the excitation energies are 
obtained for the $m=0$ low and high mode \cite{S96}
\begin{equation}
\epsilon_{TF}(m=0)_{L,H}=\hbar\omega_{\perp} \left(2+\frac{3}{2}\lambda^2\mp\frac{1}{2}
\sqrt{9\lambda^4-16\lambda^2+16}\right)^{1/2} \;\;,
\label{m0LH}
\end{equation}    
and for the surface modes of the form $\chi_m\propto r_{\perp}^{|m|} e^{im\phi}$, for which one has
\begin{equation}
\epsilon_{TF}(m)=\hbar\omega_{\perp}\sqrt{|m|} \;\;.
\label{m2}
\end{equation}
A general feature of Eq. (\ref{hydeq1}), which is explicitly reflected in the above results for
$\epsilon_{TF}$, is that the excitation energies do not depend on interaction and are proportional
to the oscillator frequencies of the harmonic potential. At finite temperature, where 
${\rm g}n_{TF}=\mu_{TF}[N_0(T)]-V_{ext}$, this fact implies that $\epsilon_{TF}$ does not depend on 
temperature either. This is an important difference with respect to the homogeneous case where
the corresponding excitations have the dispersion law $\epsilon_q=\sqrt{{\rm g}n_0/m}\;q$, and 
depend on temperature through the condensate density. The behavior exhibited in the harmonic trap
is well understood if one notes that the values of $q$ are fixed by the boundary and vary as $1/R$,
where $R$ is the size of the condensate. In the Thomas-Fermi limit, 
$R\sim\sqrt{\mu_{TF}/m\omega_{ho}^2}$ and the radius $R$ explicitly depends on the chemical potential.
On the other hand, the sound velocity is also fixed by the value of the chemical potential:
$c_B\sim\sqrt{\mu_{TF}/m}$. One finally finds that in the product $c_B q$ the chemical potential 
cancels out, so that the collective frequency is proportional to the oscillator frequency 
$\omega_{ho}$. 

Provided that finite size effects can be neglected, the explicit dependence of the collective frequency
on the interaction parameter $\eta$ as well as on the reduced temperature $t=T/T_c^0$ arises due to
quantum and thermal fluctuations which are of order ${\rm g}^2$. These fluctuations have the same physical 
origin as the corrections to the Bogoliubov speed of sound given in the homogeneous case by result 
(\ref{homEZS1}). The difference, however, is that in the case of harmonic traps, beyond mean-field effects
give corrections to a collective frequency which is fixed only by the oscillator frequency: a much better 
situation from the experimental point of view. In the following part of this section we will explicitly
calculate the effects of quantum and thermal fluctuations on the frequencies of the lowest compressional
and surface modes.  

\subsection{Perturbation scheme}

The perturbation scheme we employ for trapped systems follows the same lines as the one developed
in the homogeneous case. However, there are two important differences: first of all the 
quasiparticle states are not exact plane waves, secondly the condensate density is not fixed by 
a single parameter but depends on position.

Concerning the quasiparticle states, we make use of the local density (semiclassical) approximation
which amounts to setting \cite{GPS96,GPS97,P99}
\begin{equation}
u_i({\bf r}) = \frac{\bar{u}_i({\bf r})}{\sqrt{V}}e^{i\varphi_i({\bf r})} \;\;,\;\;\;\;
v_i({\bf r}) = \frac{\bar{v}_i({\bf r})}{\sqrt{V}}e^{i\varphi_i({\bf r})} \;\;,
\label{lda}
\end{equation}
where $V$ is a large volume contaning the system and the real functions $\bar{u}_i$, $\bar{v}_i$
satisfy the normalization condition $\bar{u}_i^2({\bf r})-\bar{v}_i^2({\bf r})=1$.
The factor $e^{i\varphi_i({\bf r})}$ represents the rapidly varying part of the functions $u_i$ and 
$v_i$, while the functions $\bar{u}_i$, $\bar{v}_i$ are assumed to be smooth functions of the 
position. The phase $\varphi_i({\bf r})$, which is also assumed to be a smooth function of 
${\bf r}$, characterizes the local impulse ${\bf p}=\hbar\nabla\varphi_i$ of the quasiparticle.
When summations over quasiparticle states are involved, these are replaced in the semiclassical 
approximation by sums over momenta, $\sum_i ...\to\sum_{\bf p} ...$, and $\bar{u}_i({\bf r})\to 
u_p({\bf r})$, $\bar{v}_i({\bf r})\to v_p({\bf r})$, where the functions $u_p({\bf r})$ and 
$v_p({\bf r})$ are given, in the region of the condensate, by the following expressions
\begin{eqnarray}
&& u_p^2({\bf r})=1+v_p^2({\bf r})=\frac{(\epsilon_p^2({\bf r})+{\rm g}^2n_{TF}^2({\bf r}))^{1/2}
+\epsilon_p({\bf r})}
{2\epsilon_p({\bf r})} \;\;,
\nonumber \\
&& u_p({\bf r})v_p({\bf r}) = - \frac{{\rm g}n_{TF}({\bf r})}{2\epsilon_p({\bf r})} \;\;,
\label{uv}
\end{eqnarray} 
where the position-dependent quasiparticle energies $\epsilon_p({\bf r})$ are given by  
\begin{equation}
\epsilon_p({\bf r}) = \left(\bigl[\epsilon_p^0+{\rm g}n_{TF}({\bf r})\bigr]^2
-\bigl[{\rm g}n_{TF}({\bf r})\bigr]^2\right)^{1/2} \;\;.
\label{bs}
\end{equation}
For each position ${\bf r}$ the above equations coincide with the Bogoliubov expressions 
(\ref{homuv}), (\ref{hombs}) with a local condensate density given by the TF density profile  
$n_{TF}({\bf r})$ defined in (\ref{nTF}). The semiclassical approximation for 
the excited states of a trapped Bose gas has been extensively used in the theoretical study 
of the thermodynamic properties of the system \cite{GPS96,GPS97,MCT97}. 
It gives a very good description of the system for temperatures $k_BT\gg\hbar\omega_{ho}$, 
but is also valid at $T=0$ if the relevant energies in the summation over excited states 
are much larger than the oscillator energy $\hbar\omega_{ho}$ \cite{DGGPS97}. 
For large systems the oscillator energy is the smallest energy scale and vanishes in the 
thermodynamic limit, as a consequence, in this limit, the semiclassical approximation becomes 
a rigorous treatment.  

The equilibrium noncondensate densities $\tilde{n}^0({\bf r})$ and $\tilde{m}^0({\bf r})$ 
are readily calculated employing the semiclassical approximation. One obtains
\begin{eqnarray}
\tilde{n}^0({\bf r}) &=& \frac{1}{V}\sum_i\left\{ [\bar{u}_i^2({\bf r})+\bar{v}_i^2({\bf r})]f_i^0 
+ \bar{v}_i^2({\bf r}) \right\} = \frac{1}{V}\sum_{\bf p}\left\{ [u_p^2({\bf r})+v_p^2({\bf r})]
f_p^0({\bf r}) + v_p^2({\bf r}) \right\} \;\;,
\nonumber\\
\tilde{m}^0({\bf r}) &=& \frac{1}{V}\sum_i \bar{u}_i({\bf r})\bar{v}_i({\bf r}) (1+2f_i^0) 
= \frac{1}{V}\sum_{\bf p} u_p({\bf r})v_p({\bf r}) \left[1+2f_p^0({\bf r})\right] \;\;,
\label{ldeq}
\end{eqnarray}
where $f_p^0({\bf r})=(e^{\epsilon_p({\bf r})/k_BT}-1)^{-1}$ is the local equilibrium quasiparticle
distribution function.

By inserting in Eq. (\ref{eqcd}) the above expressions for the noncondensate densities and 
using the renormalization (\ref{homrg}) of the coupling constant, one obtains
the following result for the profile of the condensate density valid to order ${\rm g}^2$
\begin{equation}
{\rm g}n_0({\bf r})={\rm g}n_{TF}({\bf r}) + \delta\mu - {\rm g}n_{TF}({\bf r})
[a^3n_{TF}({\bf r})]^{1/2} H[\tau({\bf r})] \;\;,
\label{eqcd1}
\end{equation}
where $H(\tau)$ is the dimensionless function (\ref{homH}) of the local reduced temperature 
$\tau({\bf r})=k_BT/{\rm g}n_{TF}({\bf r})$. In the above equation $\delta\mu=
\mu-\mu_{TF}(N_0)-2{\rm g}n_T^0$, with $n_T^0=\zeta(3/2)\lambda_T^{-3}$ as in (\ref{homdcp}),
is the shift in the chemical potential corresponding to the change in the condensate density
profile.

The application of the semiclassical approximation to the last two terms on the r.h.s. of 
Eq. (\ref{eqcf}), which 
describe the dynamic coupling to the noncondensate particles, needs a careful treatment. 
Thus, for the moment, we calculate them in terms of the $u_i$ and $v_i$ functions.
Similarly to the homogeneous case [see Eqs. (\ref{homfg})] , one must neglect in Eqs. (\ref{feq}), 
(\ref{geq}) the terms proportional to $\delta\tilde{n}$ and $\delta\tilde{m}$. 
In Fourier space, one finds for the components of the matrices $f_{ij}$, $g_{ij}$ and $g^{\ast}_{ij}$
oscillating at the frequency $\omega$
\begin{eqnarray}
f_{ij}(\omega) &=& {\rm g} \frac{f_i^0-f_j^0}{\hbar\omega+(\epsilon_i-\epsilon_j)
+i0} \int d{\bf r} \;\Phi_0
\left[(\delta\Phi_1-\delta\Phi_2)(v_iu_j^{\ast}-u_iv_j^{\ast})\right.
\nonumber\\
&+& \left. 
(\delta\Phi_1+\delta\Phi_2)(2u_iu_j^{\ast}+2v_iv_j^{\ast}+v_iu_j^{\ast}+u_iv_j^{\ast}) 
\right] \;\;,
\nonumber\\
g_{ij}(\omega) &=& {\rm g} \frac{1+f_i^0+f_j^0}{\hbar\omega-(\epsilon_i+\epsilon_j)}
\int d{\bf r} \;\Phi_0
\left[(\delta\Phi_1-\delta\Phi_2)(u_i^{\ast}u_j^{\ast}-v_i^{\ast}v_j^{\ast})\right.
\nonumber\\
&+& \left. 
(\delta\Phi_1+\delta\Phi_2)(2u_i^{\ast}v_j^{\ast}+2v_i^{\ast}u_j^{\ast}+u_i^{\ast}u_j^{\ast}
+v_i^{\ast}v_j^{\ast}) 
\right] \;\;,
\label{fgomega} \\
g_{ij}^{\ast}(\omega) &=& {\rm g} \frac{1+f_i^0+f_j^0}{\hbar\omega+(\epsilon_i+\epsilon_j)}
\int d{\bf r} \;\Phi_0
\left[(\delta\Phi_1-\delta\Phi_2)(u_iu_j-v_iv_j)\right.
\nonumber\\
&-& \left. 
(\delta\Phi_1+\delta\Phi_2)(2u_iv_j+2v_iu_j+u_iu_j+v_iv_j) 
\right] \;\;,
\nonumber
\end{eqnarray}
where we have taken $\delta\Phi({\bf r},t)=\delta\Phi_1({\bf r})\,e^{-i\omega t}$ and 
$\delta\Phi^{\ast}({\bf r},t)=\delta\Phi_2({\bf r})\,e^{-i\omega t}$. 
In Eqs. (\ref{fgomega}) we have neglected in the expressions for $g_{ij}(\omega)$ and 
$g_{ij}^{\ast}(\omega)$ the small imaginary part of the frequency $\omega$. As discussed 
in Sec. III-C, this imaginary contribution is responsible for the Beliaev damping. However,
due to discretization of levels, this damping mechanism is not effective for the low-lying 
modes we are investigating.

To order ${\rm g}^2$ the equation for the low-lying oscillations of the condensate can be 
written in the form
\begin{eqnarray}
\hbar\omega(\delta\Phi_1+\delta\Phi_2) &=& -\frac{\hbar^2\nabla^2}{2m}(\delta\Phi_1-\delta\Phi_2)
\nonumber\\
&-& 2{\rm g}\tilde{m}^0 (\delta\Phi_1-\delta\Phi_2) 
+ 2{\rm g}\sqrt{n_{TF}}\sum_{ij}(v_i^{\ast}u_j-u_i^{\ast}v_j) f_{ij}(\omega)
\nonumber\\
&+& {\rm g}\sqrt{n_{TF}}\sum_{ij}\left[ (u_iu_j-v_iv_j) g_{ij}(\omega)
- (u_i^{\ast}u_j^{\ast}-v_i^{\ast}v_j^{\ast}) g_{ij}^{\ast}(\omega) \right] \;\;,
\nonumber\\
\hbar\omega(\delta\Phi_1-\delta\Phi_2) &=& 2({\rm g}n_{TF}+\delta\mu) (\delta\Phi_1+\delta\Phi_2)
\label{dphi+-}\\
&+& 2{\rm g}n_{TF}\left[{\rm g}\frac{1}{V}\sum_{\bf p}\frac{m}{p^2}-(a^3n_{TF})^{1/2} 
H[\tau({\bf r})]\right] (\delta\Phi_1+\delta\Phi_2) 
\nonumber\\
&+& 2{\rm g}\sqrt{n_{TF}}\sum_{ij}(2u_i^{\ast}u_j+2v_i^{\ast}v_j+v_i^{\ast}u_j
+u_i^{\ast}v_j) f_{ij}(\omega)
\nonumber\\
&+& {\rm g}\sqrt{n_{TF}}\sum_{ij}\Bigl[ (2u_iv_j+2v_iu_j+u_iu_j+v_iv_j) g_{ij}(\omega) \Bigr.
\nonumber\\
&+& \Bigl. (2u_i^{\ast}v_j^{\ast}+2v_i^{\ast}u_j^{\ast}+u_i^{\ast}u_j^{\ast}+v_i^{\ast}v_j^{\ast})
g_{ij}^{\ast}(\omega) \Bigr]
\;\;.
\nonumber
\end{eqnarray}
In the above equations one can recognize the terms arising from the dynamic coupling between the 
condensate and the noncondensate component, which contain $f_{ij}(\omega)$, $g_{ij}(\omega)$ and 
$g_{ij}^{\ast}(\omega)$, the terms arising from the coupling to the static anomalous density
$\tilde{m}^0$ and from the renormalization of {\rm g}, and, finally, the terms
proportional to $\delta\mu$ and $H(\tau)$ which come from the change in the density profile 
of the condensate. The last terms have no counterpart in the homogeneous case.
In Eqs. (\ref{dphi+-}) we have neglected, as in Eqs. (\ref{hydeq}), the term proportional to 
$\nabla^2(\delta\Phi_1+\delta\Phi_2)$.

Following the analysis carried out in the homogeneous case, we write the excitation energy as
$\hbar\omega=\epsilon_{TF}+\delta E - i\gamma$. From Eqs. (\ref{dphi+-}), by treating the 
corrections to Eqs. (\ref{hydeq}) as small perturbations, one gets the result
\begin{eqnarray}
\delta E - i\gamma &=& \int d{\bf r}\;  {\rm g}n_{TF}
\left[{\rm g}\frac{1}{V}\sum_{\bf p}\frac{m}{p^2}-(a^3n_{TF})^{1/2}H[\tau({\bf r})]\right]
|\delta\Phi_1^0+\delta\Phi_2^0|^2
\nonumber\\
&-& \int d{\bf r}\; {\rm g}\tilde{m}^0 |\delta\Phi_1^0-\delta\Phi_2^0|^2
+ 4{\rm g}^2\sum_{ij} (f_i^0-f_j^0)\frac{|A_{ij}|^2}{\epsilon_{TF}+(\epsilon_i-
\epsilon_j)+i0} 
\label{encorr}\\
&+&2{\rm g}^2\sum_{ij} (1+f_i^0+f_j^0)\left( \frac{|B_{ij}|^2}{\epsilon_{TF}-
(\epsilon_i+\epsilon_j)} - \frac{|\tilde{B}_{ij}|^2}{\epsilon_{TF}+
(\epsilon_i+\epsilon_j)}\right) \;\;,    
\nonumber
\end{eqnarray}
holding for the low-lying modes with $\epsilon_{TF}\ll\mu$.  
Notice that the shift $\delta\mu$ of the chemical potential does not enter result (\ref{encorr}).
In fact, in the Thomas-Fermi limit, the excitation frequencies obtained from Eq. (\ref{hydeq1})
do not depend on the value of $\mu$. 
In Eq. (\ref{encorr}), the matrix elements $A_{ij}$, $B_{ij}$ and $\tilde{B}_{ij}$ are defined,
in analogy to the homogeneous case, as
\begin{eqnarray}
A_{ij}&=&\frac{1}{2}\int d{\bf r} \;\sqrt{n_{TF}}
\Bigl[ (\delta\Phi_1^0+\delta\Phi_2^0)(2u_iu_j^{\ast}+2v_iv_j^{\ast}+v_iu_j^{\ast}+u_iv_j^{\ast}) 
+ (\delta\Phi_1^0-\delta\Phi_2^0)(v_iu_j^{\ast}-u_iv_j^{\ast}) \Bigr] \;\;,
\nonumber\\
B_{ij}&=&\frac{1}{2}\int d{\bf r} \;\sqrt{n_{TF}}
\Bigl[ (\delta\Phi_1^0+\delta\Phi_2^0)(2u_i^{\ast}v_j^{\ast}+2v_i^{\ast}u_j^{\ast}
+u_i^{\ast}u_j^{\ast}+v_i^{\ast}v_j^{\ast}) 
+ (\delta\Phi_1^0-\delta\Phi_2^0)(u_i^{\ast}u_j^{\ast}-v_i^{\ast}v_j^{\ast}) \Bigr] \;\;,
\label{ABB}\\
\tilde{B}_{ij}&=&\frac{1}{2}\int d{\bf r} \;\sqrt{n_{TF}}
\Bigl[ (\delta\Phi_1^0+\delta\Phi_2^0)(2u_iv_j+2v_iu_j
+u_iu_j+v_iv_j) 
- (\delta\Phi_1^0-\delta\Phi_2^0)(u_iu_j-v_iv_j) \Bigr] \;\;.
\nonumber
\end{eqnarray}

Starting from Eq. (\ref{encorr}), one can study both the damping and the frequency shift of the 
low-lying modes. The calculation of the damping rates has been carried out by several authors
\cite{FSW98,BS98,GP99,RCGS99}. In Refs. \cite{FSW98,RCGS99} the damping of the $m=0$ and $m=2$ mode 
has been calculated as a function of temperature and found in good agreement with experiments 
\cite{JILA97,MIT98}. Concerning the frequency shifts, a calculation based on a method similar to
ours has been carried out in \cite{FS98}, but only for quasiclassical modes which satisfy 
the condition $\hbar\omega_{ho}\ll\epsilon_{TF}\ll\mu$. In the present work we study the 
frequency shift of the lowest-lying collective modes with $\epsilon_{TF}\sim\hbar\omega_{ho}$.
These modes have also been studied within the dielectric formalism in \cite{RCGS99}.

\subsection{Frequency shift of the collective modes}

\subsubsection{Non-resonant contribution}

Similarly to the homogeneous case, the non-resonant contribution to the frequency shift 
$\delta E$ is defined as
\begin{equation}
\delta E_{NR}=
2{\rm g}^2\sum_{ij} (1+f_i^0+f_j^0) \left( \frac{|B_{ij}|^2}{\epsilon_{TF}-\epsilon_i-\epsilon_j}
-\frac{|\tilde{B}_{ij}|^2} {\epsilon_{TF}+\epsilon_i+\epsilon_j} \right) \;\;.
\label{NR0}
\end{equation}
The matrix elements of $B$ and $\tilde{B}$ are given in (\ref{ABB}). 
By using the semiclassical approximation
(\ref{lda}) for the quasiparticle functions $u_i$ and $v_i$, one can write $\delta E_{NR}$ in the 
following form
\begin{eqnarray}
\frac{\delta E_{NR}}{\epsilon_{TF}} = - \frac{{\rm g}}{2V^2} \sum_{ij} \int d{\bf r}\, d{\bf s}\; 
e^{i{\bf s}\cdot\nabla[\varphi_i({\bf r})+\varphi_j({\bf r})]}\; 
\chi_0({\bf r})\left[ \bar{K}_{1\,ij}^{NR}({\bf r},{\bf r}+{\bf s}) +
\bar{K}_{2\,ij}^{NR}({\bf r},{\bf r}+{\bf s}) \right]  
\chi_0^{\ast}({\bf r}+{\bf s}) \;\;,
\label{NR1}
\end{eqnarray}
where the functions $\chi_0({\bf r})$ are the solutions of (\ref{hydeq1}), 
and in $e^{i{\bf s}\cdot\nabla[\varphi_i({\bf r})+\varphi_j({\bf r})]}$ we 
have neglected second derivatives of the phase $\varphi$. The smoothly varying kernels 
$\bar{K}_{1\,ij}^{NR}$ and $\bar{K}_{2\,ij}^{NR}$ are defined as 
\begin{eqnarray}
\bar{K}_{1\,ij}^{NR}({\bf r},{\bf r}^{\prime}) &=& \frac{1+f_i^0+f_j^0}{\epsilon_i+\epsilon_j}
\; \left(a_{ij}({\bf r})+\frac{2{\rm g}n_{TF}({\bf r})}{\epsilon_i+\epsilon_j}b_{ij}({\bf r})
\right) \left(a_{ij}({\bf r}^{\prime})+\frac{2{\rm g}n_{TF}({\bf r}^{\prime})}
{\epsilon_i+\epsilon_j}b_{ij}({\bf r}^{\prime})\right)  \;\;,
\nonumber\\
\bar{K}_{2\,ij}^{NR}({\bf r},{\bf r}^{\prime}) &=& \frac{1+f_i^0+f_j^0}{\epsilon_i+\epsilon_j}
\; \frac{2{\rm g}n_{TF}({\bf r})}{\epsilon_{TF}}b_{ij}({\bf r})
\frac{2{\rm g}n_{TF}({\bf r}^{\prime})}{\epsilon_{TF}}b_{ij}({\bf r}^{\prime})\;\;,
\label{ker}
\end{eqnarray}
where we have introduced the matrices
\begin{eqnarray}
a_{ij}({\bf r}) &=& 2\bar{u}_i({\bf r})\bar{v}_j({\bf r})+2\bar{v}_i({\bf r})\bar{u}_j({\bf r})
+\bar{u}_i({\bf r})\bar{u}_j({\bf r})+\bar{v}_i({\bf r})\bar{v}_j({\bf r}) \;\;,
\nonumber\\
b_{ij}({\bf r}) &=& \bar{u}_i({\bf r})\bar{u}_j({\bf r})-\bar{v}_i({\bf r})\bar{v}_j({\bf r}) \;\;.
\label{ab}
\end{eqnarray}
In the limit $\epsilon_{TF}\ll\mu$ we can use the following gradient expansion
\begin{eqnarray}
\chi_0({\bf r})\bar{K}_{1\,ij}^{NR}({\bf r},{\bf r}+{\bf s})\chi_0^{\ast}({\bf r}+{\bf s}) \simeq
\bar{K}_{1\,ij}^{NR}({\bf r},{\bf r})\; |\chi_0({\bf r})|^2 \;\;,
\label{K1}
\end{eqnarray}
and
\begin{eqnarray}
\chi_0({\bf r})\bar{K}_{2\,ij}^{NR}({\bf r},{\bf r}+{\bf s})\chi_0^{\ast}({\bf r}+{\bf s}) &\simeq&
\bar{K}_{2\,ij}^{NR}({\bf r},{\bf r}) \Bigl[ |\chi_0({\bf r})|^2 
+ \frac{1}{2}\chi_0({\bf r})\left({\bf s}\cdot\nabla\right)^2\chi_0^{\ast}({\bf r}) \Bigr] 
\nonumber\\ 
&+& \frac{1}{2} \chi_0({\bf r})\left({\bf s}\cdot\nabla\right)\bar{K}_{2\,ij}^{NR}({\bf r},{\bf r})
\left({\bf s}\cdot\nabla\right)
\chi_0^{\ast}({\bf r}) \;\;.
\label{K2}
\end{eqnarray}
Since the kernel $\bar{K}_1^{NR}$ is already zeroth order in $\epsilon_{TF}/\mu$, we can neglect 
higher order terms in the expansion (\ref{K1}). On the contrary, $\bar{K}_2^{NR}$ is of order 
$(\mu/\epsilon_{TF})^2$ and we need the expansion (\ref{K2}) to second order in the displacement
${\bf s}$. Notice also that terms in the expansion (\ref{K2}) containing odd powers of ${\bf s}$  
vanish in (\ref{NR1}) due to geometry. Moreover, we have neglected in (\ref{K2}) second derivatives 
of the slowly varying functions $\bar{u}_i$, $\bar{v}_i$. 
By the replacement $\sum_i\to\sum_{\bf p}$, and after integration by parts, one gets the result
\begin{eqnarray}
\frac{\delta E_{NR}}{\epsilon_{TF}} &=& - \frac{{\rm g}}{2V^2} \int d{\bf r}\, d{\bf s} 
\, |\chi_0({\bf r})|^2 \sum_{{\bf p}{\bf q}} e^{-i{\bf q}\cdot{\bf s}/\hbar} 
\left[ \bar{K}_{1\,pk}^{NR}({\bf r},{\bf r}) +
\bar{K}_{2\,pk}^{NR}({\bf r},{\bf r}) \right] 
\nonumber\\
&+& \frac{{\rm g}}{12V^2} \int d{\bf r}\, d{\bf s} 
\, |\nabla\chi_0({\bf r})|^2 \sum_{{\bf p}{\bf q}} s^2 e^{-i{\bf q}\cdot{\bf s}/\hbar} 
\bar{K}_{2\,pk}^{NR}({\bf r},{\bf r}) \;\;, 
\label{NR2}
\end{eqnarray}
where ${\bf k}={\bf q}+{\bf p}$. In the first term on the r.h.s. of (\ref{NR2}) the integration 
over ${\bf s}$ gives $\delta_{{\bf q}{\bf 0}}$, while in the second term one writes 
$s^2e^{-i{\bf q}\cdot{\bf s}/\hbar}=-\hbar^2\nabla_{\bf q}^2
e^{-i{\bf q}\cdot{\bf s}/\hbar}$ and integrates by parts over ${\bf q}$. 
After some algebra one gets the result
\begin{eqnarray}
\frac{\delta E_{NR}}{\epsilon_{TF}} &=& - \int d{\bf r}\,
{\rm g}\frac{1}{V}\sum_{\bf p}(1+2f_p^0) 
\left[ |\chi_0|^2 \frac{(\epsilon_p^0)^2}{4\epsilon_p^3} - |\nabla\chi_0|^2 
\frac{\hbar^2{\rm g}^2n_{TF}^2}{m\epsilon_{TF}^2}\frac{\epsilon_p^0}{6\epsilon_p^3} \right]
\nonumber\\
&-& \int d{\bf r}\,|\chi_0|^2 \frac{{\rm g}^2n_{TF}^2}{\epsilon_{TF}^2}
{\rm g}\frac{1}{V}\sum_{\bf p}\frac{1+2f_p^0}{\epsilon_p}
\label{NR3}\\
&-& \frac{1}{6} \int d{\bf r}\,|\nabla\chi_0|^2 \frac{\hbar^2{\rm g}^2n_{TF}^2}{\epsilon_{TF}^2}
\frac{1}{V}\sum_{\bf q}\int d{\bf s}\, e^{-i{\bf q}\cdot{\bf s}/\hbar}\nabla_{\bf q}^2 \;{\rm g}
\frac{1}{V}\sum_{\bf p}\frac{f_p^0}{\epsilon_{|{\bf q}+{\bf p}|}} \;\;.
\nonumber
\end{eqnarray}
In the above equation $\epsilon_p=\epsilon_p({\bf r})$ and $f_p^0=f_p^0({\bf r})$, according to 
(\ref{bs}). We notice that in the homogeneous limit, where $\chi_0({\bf r})=
e^{i{\bf q}\cdot{\bf r}/\hbar}/\sqrt{V}$ and $\epsilon_{TF}=c_B q$, the first two terms 
on the r.h.s. of (\ref{NR3}) coincide with the corresponding terms in (\ref{homENR1}).

\subsubsection{Resonant contribution}

The resonant contribution to $\delta E$ is defined as
\begin{equation} 
\delta E_{R}=4{\rm g}^2\sum_{ij}(f_i^0-f_j^0) \frac{|A_{ij}|^2} {\epsilon_{TF}+\epsilon_i-
\epsilon_j} \;\;,
\label{R0}
\end{equation}
where the matrix elements $A_{ij}$ are given in (\ref{ABB}). 
Following the method used in the analysis of the non-resonant terms one has
\begin{eqnarray}
\frac{\delta E_{R}}{\epsilon_{TF}} = \frac{{\rm g}}{2V^2} \sum_{ij} \int d{\bf r}\, d{\bf s}\; 
e^{-i{\bf s}\cdot\nabla[\varphi_i({\bf r})-\varphi_j({\bf r})]}\; 
\chi_0({\bf r})\left[ \bar{K}_{1\,ij}^{R}({\bf r},{\bf r}+{\bf s}) +
\bar{K}_{2\,ij}^{R}({\bf r},{\bf r}+{\bf s}) \right]  
\chi_0^{\ast}({\bf r}+{\bf s}) \;\;.
\label{R1}
\end{eqnarray}
The kernels in the above equation are defined by 
\begin{eqnarray}
\bar{K}_{1\,ij}^{R}({\bf r},{\bf r}^{\prime}) &=& \frac{f_i^0-f_j^0}{\epsilon_{TF}+\epsilon_i
-\epsilon_j}
\; \left(c_{ij}({\bf r})-\frac{2{\rm g}n_{TF}({\bf r})}{\epsilon_i-\epsilon_j}d_{ij}({\bf r})
\right) \left(c_{ij}({\bf r}^{\prime})-\frac{2{\rm g}n_{TF}({\bf r}^{\prime})}
{\epsilon_i-\epsilon_j}d_{ij}({\bf r}^{\prime})\right)  \;\;,
\nonumber\\
\bar{K}_{2\,ij}^{R}({\bf r},{\bf r}^{\prime}) &=& \frac{f_i^0-f_j^0}{\epsilon_i-\epsilon_j}
\; \frac{2{\rm g}n_{TF}({\bf r})}{\epsilon_{TF}}d_{ij}({\bf r})
\frac{2{\rm g}n_{TF}({\bf r}^{\prime})}{\epsilon_{TF}}d_{ij}({\bf r}^{\prime})\;\;,
\label{ker1}
\end{eqnarray}
where we have introduced the matrices
\begin{eqnarray}
c_{ij}({\bf r}) &=& 2\bar{u}_i({\bf r})\bar{u}_j({\bf r})+2\bar{v}_i({\bf r})\bar{v}_j({\bf r})
+\bar{v}_i({\bf r})\bar{u}_j({\bf r})+\bar{u}_i({\bf r})\bar{v}_j({\bf r}) \;\;,
\nonumber\\
d_{ij}({\bf r}) &=& \bar{v}_i({\bf r})\bar{u}_j({\bf r})-\bar{u}_i({\bf r})\bar{v}_j({\bf r}) \;\;.
\label{cd}
\end{eqnarray}
In the limit $\epsilon_{TF}\ll\mu$ the term in Eq. (\ref{R1}) containing the kernel 
$\bar{K}_{2\,ij}^{R}$ can be treated using the gradient expansion (\ref{K2}). One gets thus
\begin{eqnarray}
\frac{{\rm g}}{2V^2} \sum_{ij} \int d{\bf r}\, d{\bf s}\; &&
e^{-i{\bf s}\cdot\nabla[\varphi_i({\bf r})-\varphi_j({\bf r})]}\; \chi_0({\bf r})  
\bar{K}_{2\,ij}^{R}({\bf r},{\bf r}+{\bf s}) \chi_0^{\ast}({\bf r}+{\bf s}) =
\nonumber\\
&& - \int d{\bf r}\, |\nabla\chi_0|^2 \frac{\hbar^2{\rm g}^2n_{TF}^2}{m\epsilon_{TF}^2} 
{\rm g}\frac{1}{V}\sum_{\bf p}\frac{\partial f_p^0}{\partial\epsilon_p}
\frac{\epsilon_p^0}{3\epsilon_p^2}
\label{R2}\\
&& + \frac{1}{6} \int d{\bf r}\,|\nabla\chi_0|^2 \frac{\hbar^2{\rm g}^2n_{TF}^2}{\epsilon_{TF}^2}
\frac{1}{V}\sum_{\bf q}\int d{\bf s}\, e^{-i{\bf q}\cdot{\bf s}/\hbar}\nabla_{\bf q}^2 \;
{\rm g}\frac{1}{V}\sum_{\bf p}\frac{f_p^0}{\epsilon_{|{\bf q}+{\bf p}|}} \;\;.
\nonumber
\end{eqnarray} 
In the homogeneous limit, the first term on the r.h.s. of (\ref{R2}) coincides with the second term
in the square bracket on the r.h.s. of (\ref{homER1}). 
Moreover, the last term on the r.h.s. of (\ref{NR3}) and (\ref{R2})
are equal and opposite in sign, and cancel out in the sum $\delta E_{NR}+\delta E_{R}$. A similar 
cancellation is also present in the homogeneous case [see Eqs. (\ref{homENR1}) and (\ref{homER1})]. 

The contribution to $\delta E_{R}$ arising from $\bar{K}_{1\,ij}^{R}$ is more delicate.
In fact, all terms in the gradient expansion give contributions which are of the same order 
in the limit $\epsilon_{TF}\ll\mu$. Hovever, if we restrict ourselves to modes for which 
$\nabla^2\chi_0=const$, the expansion (\ref{K2}) is still appropriate. In fact, higher 
order terms in (\ref{K2}) contain derivatives of $\nabla^2\chi_0$ and vanish for modes with 
constant Laplacian. To this class of modes belong, for example, all surface modes, for which 
$\nabla^2\chi_0=0$, and the lowest breathing modes.     
For the above mentioned term one finds 
\begin{eqnarray}
\frac{{\rm g}}{2V^2} \sum_{ij} && \int d{\bf r}\, d{\bf s}\; 
e^{-i{\bf s}\cdot\nabla[\varphi_i({\bf r})-\varphi_j({\bf r})]}\; 
\chi_0({\bf r})\bar{K}_{1\,ij}^{R}({\bf r},{\bf r}+{\bf s}) \chi_0^{\ast}({\bf r}+{\bf s}) =
\nonumber\\
&& - \frac{1}{12} \int d{\bf r}\, d{\bf s}\, {\rm g}\frac{1}{V^2}\sum_{{\bf p}{\bf q}} 
e^{-i{\bf q}\cdot{\bf s}/\hbar} \hbar^2\nabla_{\bf q}^2 
\left[ \chi_0({\bf r})\nabla^2\chi_0^{\ast}({\bf r}) \bar{K}_{1\,pk}^{R}({\bf r},{\bf r}) 
+ \chi_0({\bf r})\nabla\chi_0^{\ast}({\bf r}) \cdot 
\nabla \bar{K}_{1\,pk}^{R}({\bf r},{\bf r}) \right] \;\;.
\label{R3}
\end{eqnarray}
The Laplacian in momentum space can be easily calculated once the low-${\bf q}$
behavior of $\bar{K}_{1\,pk}^{R}({\bf r},{\bf r})$ has been obtained. A straightforward
calculation yields:
\begin{equation}
{\rm g}\frac{1}{V}\sum_{\bf p} \bar{K}_{1\,p|{\bf q}+{\bf p}|}^{R} \to 
\frac{2 q^2}{3m\epsilon_{TF}^2} {\rm g}\frac{1}{V}\sum_{\bf p} \left(-\frac{\partial f_p^0}
{\partial\epsilon_p}\right) \epsilon_p^0 \left(1+\frac{\epsilon_p^0}{\epsilon_p}
\frac{\partial\epsilon_p}{\partial\epsilon_p^0}\right)^2 \;\;.
\label{Lqex}
\end{equation}
After some algebra, one obtaines the following result for the contribution to $\delta E_{R}$ 
\begin{eqnarray}
\frac{{\rm g}}{2V^2} \sum_{ij} \int && d{\bf r}\, d{\bf s}\; 
e^{-i{\bf s}\cdot\nabla[\varphi_i({\bf r})-\varphi_j({\bf r})]}\;
\chi_0({\bf r})   
\bar{K}_{1\,ij}^{R}({\bf r},{\bf r}+{\bf s}) \chi_0^{\ast}({\bf r}+{\bf s}) =
\frac{\sqrt{32}}{3\sqrt{\pi}} \frac{\hbar^2{\rm g}}{m\epsilon_{TF}^2}
\nonumber\\
&& \times \int d{\bf r}\, |\nabla\chi_0|^2 (an_{TF})^{3/2}
 \int dx \frac{\tau x}{(e^{x/2}-e^{-x/2})^2}
\left[ \frac{(u-1)^{3/2}}{u} \left(\frac{2u+1}{u+1}\right)^2 - 4\sqrt{\tau x}\right] \;\;,
\label{R4}
\end{eqnarray}
where $u=\sqrt{1+\tau^2({\bf r})x^2}$ and $\tau({\bf r})=k_BT/{\rm g}n_{TF}({\bf r})$. 
We stress that the above contribution to the frequency shift 
is peculiar of collective modes with constant Laplacian of harmonically trapped systems in the 
thermodynamic limit. In the homogeneous case, where $\chi_0\propto e^{i{\bf q}\cdot{\bf r}}$ and 
$\nabla^2\chi_0\neq const$, the contribution of this term is different and is given by the first 
term in the square bracket on the r.h.s. of (\ref{homER1}).   

\subsubsection{Results}

We are now in a position to calculate the shift $\delta E$ to order ${\rm g}^2$, by summing the 
various contributions to the real part of Eq. (\ref{encorr}). 
One gets the relevant result
\begin{eqnarray}
\frac{\delta E}{\epsilon_{TF}} &=& - \frac{4}{3\sqrt{\pi}}\frac{{\rm g}}{m\omega_{TF}^2}
\int d{\bf r}\, (an_{TF})^{3/2}\left(\chi_0\nabla^2\chi_0^{\ast} + \chi_0^{\ast}\nabla^2\chi_0\right)
\nonumber\\
&+& \int d{\bf r}\, (a^3n_{TF})^{1/2} |\chi_0|^2 G_1[\tau({\bf r})]
- \frac{{\rm g}}{m\omega_{TF}^2} \int d{\bf r}\, (an_{TF})^{3/2} |\nabla\chi_0|^2 G_2[\tau({\bf r})]
\;\;,
\label{DE}
\end{eqnarray}
where $\omega_{TF}=\epsilon_{TF}/\hbar$, and the functions $G_1(\tau)$, $G_2(\tau)$ of the local 
reduced temperature $\tau({\bf r})=k_BT/{\rm g}n_{TF}({\bf r})$ are defined as follows
\begin{equation}
G_1(\tau) = \frac{\sqrt{32}}{\sqrt{\pi}} \tau \int_0^{\infty} dx \frac{1}{e^x-1} 
\left( \sqrt{\tau x}-\frac{\sqrt{u-1}}{u}\,\frac{u^2+u-1}{u+1} \right) 
\;\;,
\label{G1}
\end{equation}
and 
\begin{eqnarray}
G_2(\tau) = \frac{\sqrt{32}}{3\sqrt{\pi}} \tau && \Biggl\{ \int_0^{\infty} dx 
\frac{x}{(e^{x/2}-e^{-x/2})^2}
\left[ 4\sqrt{\tau x} - \frac{\sqrt{u-1}}{u(u+1)} - \frac{(u-1)^{3/2}}{u} 
\left(\frac{2u+1}{u+1}\right)^2 \right] \Biggr.
\nonumber\\
&& - \Biggl. \int_0^{\infty} dx \frac{1}{e^x-1}\, \frac{\sqrt{u-1}}{u(u+1)} \Biggr\} \;\;,
\label{G2}
\end{eqnarray}
where, as usual, $u=\sqrt{1+\tau^2x^2}$.
The functions $G_1(\tau)$ and $G_2(\tau)$ are plotted in Fig. 4. Both are positive for any value of 
$\tau$. As a consequence, $G_1(\tau)$ gives an upward shift of the excitation frequency, while 
$G_2(\tau)$ gives a downward shift. The above result holds for collective modes which do not excite 
the center of mass degrees of freedom. In fact, as discussed in Sec. II, the theoretical approach 
developed in the present work does not describe the center of mass motion and, in particular, the 
dipole mode. 
  
At $T=0$, both $G_1$ and $G_2$ are zero and one is left with the result
\begin{equation}
\frac{\delta E}{\epsilon_{TF}} = - \frac{4}{3\sqrt{\pi}}\frac{{\rm g}}{m\omega_{TF}^2}
\int d{\bf r}\, (an_{TF})^{3/2}\left(\chi_0\nabla^2\chi_0^{\ast} + \chi_0^{\ast}\nabla^2\chi_0\right)
\;\;,
\label{DE0}
\end{equation}
which coincides with the findings of Ref. \cite{PS98}, obtained from the hydrodynamic theory of 
superfluids, and of Ref. \cite{BP99}. Notice that only non-resonant terms give contribution 
at $T=0$, as a consequence result (\ref{DE0}) holds in general for collective modes with 
$\epsilon_{TF}\ll\mu$, and is not restricted to modes which have constant Laplacian. 
For the monopole (breathing) mode in a spherically symmetric 
trap ($\lambda=1$), characterized by the frequency $\omega_{M}=\sqrt{5}\omega_{ho}$, one has from 
Eq. (\ref{DE0}) the fractional shift \cite{PS98,BP99}
\begin{equation}
\frac{\delta\omega_M}{\omega_M} = \frac{21\sqrt{2}}{320\zeta(3)} \eta^3 \;\;,
\label{M0}
\end{equation}
expressed in terms of the parameter $\eta$.
For the $m=0$ modes in an axially symmetric trap, which have excitation frequency given by 
(\ref{m0LH}), one finds the result \cite{PS98,BP99}
\begin{equation}
\frac{\delta\omega_{m=0}}{\omega_{m=0}} = \frac{21\sqrt{2}}{320\zeta(3)} \eta^3 f_{\pm}(\lambda) \;\;,
\label{m=00}
\end{equation} 
where 
\begin{equation}
f_{\pm}({\lambda})=\frac{1}{2} \pm \frac{8+\lambda^2}{6\sqrt{9\lambda^4-16\lambda^2+16}} \;\;,
\label{fpm}
\end{equation}
and the index $\pm$ refers to the high ($+$) and low ($-$) $m=0$ mode.
As discussed in Ref. \cite{PS98}, these frequency shifts are very small. For $\eta=0.4$, which is a 
typical value for the interaction parameter in experiments, one gets a fractional shift of 
the order of 0.5 \%.

At finite temperature the terms involving the $G_1$ and $G_2$ functions contribute to the frequency
shift, and, differently from the $T=0$ case, also surface excitations with $\nabla^2\chi_0=0$ are 
affected by the correction (\ref{DE}).
We consider first spherically symmetric traps. The monopole oscillation has the form 
$\chi_M\propto (r^2-3R^2/5)$, where $R=\sqrt{2\mu_{TF}(N_0)/m\omega_{ho}^2}$ is the condensate radius.
The temperature dependence of the fractional shift $\delta\omega_M/\omega_M$ is given by the 
equation
\begin{eqnarray}
\frac{\delta\omega_M}{\omega_M} = \frac{21\sqrt{2}}{320\zeta(3)} \eta^3 
\left(\frac{N_0}{N}\right)^{1/5} && \left[ 1+\frac{16}{9\sqrt{\pi}}\int_0^1 dx\, x^{1/2}\sqrt{1-x}
\,(2-5x)^2\, G_1[\tau(x)] \right.
\nonumber\\
&& \left. - \frac{160}{9\sqrt{\pi}}\int_0^1 dx\, x^{3/2} (1-x)^{3/2} \,G_2[\tau(x)] \right]
\;\;,
\label{MT}
\end{eqnarray}  
where $N_0/N$ is the equilibrium value of the condensate fraction, which is fixed by the parameter 
$\eta$ and the reduced temperature $t=T/T_c^0$. The argument $\tau(x)$ of the $G_1$ and $G_2$ 
functions is given by the expression $\tau(x)=[(N_0/N)^{-2/5}t/\eta] 1/x$.  
In Fig. 6 the monopole frequency shift (\ref{MT}) is shown as a function of the reduced temperature 
$t$ for the value $\eta=0.4$ of the interaction parameter.
Already at relatively low temperatures, $t\simeq 0.3$, the monopole frequency is found to be about 
1 \% smaller than $\omega_M=\sqrt{5}\omega_{ho}$. In fact, even for such low temperatures, 
the local reduced temperature is large at the boundary of the condensate, as $\tau(x)\gg 1$ if 
$x\to 0$, and the contribution of this region dominates the shift (\ref{MT}).     
If $\eta\ll t$, one can approximate the functions $G_1$ and $G_2$ in (\ref{MT}) with their asymptotic 
behavior for $\tau\gg1$. One has 
\begin{eqnarray}
G_1(\tau) &\to& G_1(\infty) \tau \;\;,
\nonumber\\
G_2(\tau) &\to& G_2(\infty) \tau \;\;,
\label{G12infty}
\end{eqnarray}
where $G_1(\infty)=5\sqrt{\pi}$ and 
\begin{equation}
G_2(\infty) = \frac{4\sqrt{2}}{3\sqrt{\pi}} \int_0^1 dx\,
\left[ \frac{8+5x-x^2}{\sqrt{x}\sqrt{1-x}(1+x)^3} + \frac{4}{x^{3/2}(1-x^2)^{3/4}} 
\left( 1-\frac{(1-x)^{1/4}}{(1+x)^{1/4}} \right)\right]
\simeq 22.
\label{G2inf}
\end{equation} 
In this case the monopole shift is given by
\begin{equation}
\frac{\delta\omega_M}{\omega_M} = \frac{7\sqrt{2\pi}}{960\zeta(3)}\;\frac{\eta^2 t}{(1-t^3)^{1/5}}
\left[17G_1(\infty)-10G_2(\infty)\right] \simeq - 1.1 \frac{\eta^2 t}{(1-t^3)^{1/5}} \;\;,
\label{MHT}
\end{equation}
where for the condensate fraction we have used the ideal gas law $N_0/N=1-t^3$.
Result (\ref{MHT}) gives a reasonably good approximation to the frequency shift $\delta\omega_M$ also 
when $\eta\sim t$, for example, for $\eta=0.4$ and $t=0.8$ Eq. (\ref{MHT}) gives 
$\delta\omega_M/\omega_M\simeq -0.16$ and the calculation based on Eq. (\ref{MT}) gives $-0.11$.
  
In a surface mode the oscillation of the condensate has the form $\chi_{lm}\propto 
r^l Y_{lm}(\theta,\phi)$, and the excitation frequency is given by $\omega_l=\sqrt{l}\omega_{ho}$.
For these modes one finds the following fractional shift
\begin{eqnarray}
\frac{\delta\omega_l}{\omega_l} = \frac{\sqrt{2}(2l+3)}{15\sqrt{\pi}\zeta(3)} \eta^3 
\left(\frac{N_0}{N}\right)^{1/5} && \left[ \int_0^1 dx\, x^{1/2} (1-x)^{l+1/2}
\, G_1[\tau(x)] \right.
\nonumber\\ 
&& \left. - (l+1/2) \int_0^1 dx\, x^{3/2} (1-x)^{l-1/2} \,G_2[\tau(x)] \right]
\;\;.
\label{lmT}
\end{eqnarray}
In the limit $\eta\ll t$ one gets, by using (\ref{G12infty}), the following result
\begin{equation}
\frac{\delta\omega_l}{\omega_l} = \frac{\sqrt{2}(2l+3)}{30\zeta(3)(l+1)}\,
\frac{(l+\frac{1}{2})\Gamma(l+\frac{1}{2})}{l\,\Gamma(l)}\; 
\frac{\eta^2 t}{(1-t^3)^{1/5}} \left[2G_1(\infty)-
G_2(\infty) \right] \;\;.
\label{lmHT}
\end{equation}

In the case of axially symmetric traps, the oscillations with $m=0$ symmetry have the form
$\chi_{m=0}\propto -2\mu_{TF}(s^2-2)/[ms^2\omega_{\perp}^2 + r_{\perp}^2 + (s^2-4)z^2]$, where
$s=\omega_{m=0}/\omega_{\perp}$ and $\omega_{m=0}$ is given by Eq. (\ref{m0LH}) for the low and high
mode. After some algebra one gets the result
\begin{eqnarray}
\frac{\delta\omega_{m=0}}{\omega_{m=0}} = \frac{21\sqrt{2}}{320\zeta(3)} \eta^3 &&
\left(\frac{N_0}{N}\right)^{1/5}  
 \left\{ f_{\pm}(\lambda) - \frac{160}{9\sqrt{\pi}}\int_0^1 dx\, x^{3/2} (1-x)^{3/2} 
\,G_2[\tau(x)]  \right.
\\
\label{m=0T}
&& \left. + \frac{16}{9\sqrt{\pi}}\int_0^1 dx\, x^{1/2}\sqrt{1-x}
\,\biggl[4(1-x)^2+3f_\pm(\lambda)(7x^2-4x)\biggr]\, G_1[\tau(x)]  \right\}
\;\;,
\nonumber
\end{eqnarray}
where $f_\pm(\lambda)$ is defined in (\ref{fpm}). In the limit $\eta\ll t$ the above result 
reduces to
\begin{eqnarray}
\frac{\delta\omega_{m=0}}{\omega_{m=0}} = 
\frac{7\sqrt{2\pi}}{960\zeta(3)}\;\frac{\eta^2 t}{(1-t^3)^{1/5}}
\left\{ [20-3f_\pm(\lambda)] G_1(\infty)-10G_2(\infty)\right\} \;\;.
\label{m=0HT}
\end{eqnarray} 
On the contrary, surface excitations of the form $\chi_{m}\propto r_{\perp}^{|m|} e^{im\phi}$ 
and with excitation energy $\omega_m=\sqrt{|m|}\omega_{\perp}$, exhibit the fractional shift 
(\ref{lmT}) with $l$ replaced by $|m|$.
 
In Fig. 5 (Fig. 6) we show the fractional shift of the mode $m=0$ low (high) as a function of the 
reduced temperature $t=T/T_c^0$ and for the value $\eta=0.4$ of the interaction parameter.
We notice that in the case of the $m=0$ high mode (Fig. 6), the size of the fractional shift is 
maximum for 
spherically symmetric traps ($\lambda=1$) and is minimum for disk-shaped traps ($\lambda\gg 1$).
On the contrary, for the $m=0$ low mode (Fig. 5), $|\delta\omega/\omega|$ is maximum for 
$\lambda\gg 1$, while 
it is minimum in the $\lambda=1$ case. However, by changing the geometry of the trap, the curve of 
the fractional shift remains qualitatively the same, and at intermediate temperatures, 
$T\sim 0.5 T_c^0$, one finds downward shifts ranging from 1 to 4\% for both the mode $m=0$ low and 
high. In Fig. 7 we show the results for the surface modes $m=2$ and $m=4$ for the same value, 
$\eta=0.4$, of the interaction parameter. In the case of surface modes the fractional shift is 
independent of the deformation parameter $\lambda$ and we find that the size of the shift 
increases by increasing $m$. An explanation of this behavior can be found in the fact that modes 
with higher $m$ are more localized at the surface of the condensate where $k_BT\gg 
{\rm g}n_0({\bf r})$, being ${\rm g}n_0({\bf r})$ the local chemical potential. Thus, thermal effects
are more pronounced for such modes.

Experiments on the temperature dependence of the collective modes have been carried out both at JILA 
\cite{JILA97} and MIT \cite{MIT98}. The JILA group has measured, as a function of temperature, the 
frequency of the $m=2$ and $m=0$ low modes \cite{JILA97}. However, in these experiments, the number 
of trapped particles is about 10$^4$ and beyond Thomas-Fermi effects are expected to play a 
significant role. Nevertheless, our results for the fractional shift of the $m=2$ mode in disk-shaped 
geometries, shown in Fig. 7, both qualitativelly and quantitativelly agree with the observed behavior.
In the case of the $m=0$ low mode other effects, not included in the present analysis, might be
responsible for the features observed in the experiment. The frequency of collective excitations in 
the Thomas-Fermi regime has been measured by the MIT group for the $m=0$ low mode in a cigar-shaped 
trap \cite{MIT98}. In Fig. 8 we show the comparison between the experimental results and our 
theoretical prediction. The calculation has been carried out with the value $\eta=0.4$ of the 
interaction parameter, which is close to the experimental conditions of \cite{MIT98}. In Fig. 8,
the experimental data have been plotted as a function of the reduced temperature $T/T_c^0$ 
\cite{DSK}. This is possible only for temperatures above 0.5 $\mu$K, as lower temperatures were not 
measurable in \cite{MIT98}.

\subsection{Hydrodynamic equations at $T=0$}

At zero temperature, superfluid systems are described by the equations of hydrodynamics
(for a general discussion see the book \cite{K65}). These equations involve the total 
density $n$ of the system and the superfluid velocity ${\bf v}_s$, which is related to 
the gradient of the phase of the order parameter. The hydrodynamic picture has been 
successfully employed in \cite{S96} to obtain the frequencies of the collective modes
in the Thomas-Fermi regime and later in \cite{PS98} to calculate the corrections to these 
frequencies due to beyond mean-field effects. We have already verified [see Eq. (\ref{DE0})]
that our perturbation scheme reproduces at $T=0$ the results obtained from hydrodynamic theory. 
However, since we start from dynamic equations written in terms of the condensate wavefunction and the 
noncondensate density, it is important to understand whether these equations reduce at zero 
temperature to the hydrodynamics of superfluids.

At the level of Gross-Pitaevskii theory the analogy is straightforward \cite{S96,WG96}. By writing 
the condensate wavefunction in terms of a modulus and a phase $\Phi({\bf r},t)=
\sqrt{n_0({\bf r},t)}e^{i\varphi({\bf r},t)}$, one has the following identifications 
\begin{eqnarray}
\delta n_0({\bf r},t) &=& \Phi_0({\bf r})\left[\delta\Phi({\bf r},t)+
\delta\Phi^{\ast}({\bf r},t)\right] 
\;\;,\nonumber\\
i\delta\varphi({\bf r},t) &=& \frac{1}{2\Phi_0({\bf r})}
\left[\delta\Phi({\bf r},t)-\delta\Phi^{\ast}({\bf r},t)\right] \;\;,
\label{hyddv}
\end{eqnarray}
between the fluctuations of $n_0$ and $\varphi$ and the fluctuations of the order parameter.
The coupled equations (\ref{hydeq}), holding in the Thomas-Fermi regime, are then equivalent to
\begin{eqnarray}
&& \frac{\partial\delta n_0}{\partial t} + \nabla\cdot\left(n_{TF}{\bf v}_s\right) = 0 \;\;,
\nonumber\\
&& m\frac{\partial{\bf v}_s}{\partial t} + {\rm g}\nabla\delta n_0 = 0 \;\;,
\label{shydeq}
\end{eqnarray}
where ${\bf v}_s=\hbar\nabla\varphi/m$ is the superfluid velocity. 
At $T=0$, if one neglects quantum depletion, the condensate density coincides with the 
total density and Eqs. (\ref{shydeq}) coincide with the linearized hydrodynamic equations.
The former of Eqs. (\ref{shydeq}) corresponds to the equation of continuity and the latter 
to Euler equation with the pressure $P$ fixed by $\partial P/\partial n_0={\rm g}n_{TF}$.

Beyond Gross-Pitaevskii theory one must replace Eqs. (\ref{hydeq}) by (\ref{dphi+-}), which include 
the corrections to order ${\rm g}^2$. At $T=0$ these equations reduce to     
\begin{eqnarray}
\hbar\omega(\delta\Phi_1+\delta\Phi_2) &=& -\frac{\hbar^2\nabla^2}{2m}(\delta\Phi_1-\delta\Phi_2)
- 2{\rm g}\tilde{m}^0 (\delta\Phi_1-\delta\Phi_2)
\nonumber\\
&+& {\rm g}\sqrt{n_{TF}}\sum_{ij}\left[ (u_iu_j-v_iv_j) g_{ij}(\omega)
- (u_i^{\ast}u_j^{\ast}-v_i^{\ast}v_j^{\ast}) g_{ij}^{\ast}(\omega) \right] \;\;,
\nonumber\\
\hbar\omega(\delta\Phi_1-\delta\Phi_2) &=& 2({\rm g}n_{TF}+\delta\mu) (\delta\Phi_1+\delta\Phi_2)
\label{dphi+-0}\\
&+& 2{\rm g}n_{TF}\left[{\rm g}\frac{1}{V}\sum_{\bf p}\frac{m}{p^2}-\frac{40}{3\sqrt{\pi}}
(a^3n_{TF})^{1/2}\right] (\delta\Phi_1+\delta\Phi_2) 
\nonumber\\
&+& {\rm g}\sqrt{n_{TF}}\sum_{ij}\Bigl[ (2u_iv_j+2v_iu_j+u_iu_j+v_iv_j) g_{ij}(\omega) \Bigr.
\nonumber\\
&+& \Bigl. (2u_i^{\ast}v_j^{\ast}+2v_i^{\ast}u_j^{\ast}+u_i^{\ast}u_j^{\ast}+v_i^{\ast}v_j^{\ast})
g_{ij}^{\ast}(\omega) \Bigr]
\;\;,
\nonumber
\end{eqnarray}
where $\delta\mu=\mu-\mu_{TF}(N_0)$ is the change in the chemical potential 
[see Eq. (\ref{eqcd1})] and the matrices $g_{ij}(\omega)$ and $g_{ij}^\ast(\omega)$ are given
in (\ref{fgomega}) with $f_i^0=f_j^0=0$. By using the semiclassical approximation (\ref{lda}) 
for the quasiparticle states, the above equations are written in terms of the variables 
$\varphi$ and $n_0$ as
\begin{eqnarray}
- i\hbar\omega\; \delta n_0({\bf r}) &=& -\hbar\nabla\cdot 
\left[ n_0({\bf r}){\bf v}_s({\bf r})\right]
- 4{\rm g}n_{TF}({\bf r})
\tilde{m}^0({\bf r})\delta\varphi({\bf r}) 
\nonumber\\
&+& \frac{2{\rm g}^2}{V^2}\sum_{ij} \frac{n_{TF}({\bf r})b_{ij}({\bf r})}{\epsilon_i+\epsilon_j}
\int d{\bf s}\; e^{i{\bf s}\cdot\nabla[\varphi_i({\bf r})+\varphi_j({\bf r})]}\;
\label{hydro0}\\
&\times& \left[\frac{i\hbar\omega\delta n_0({\bf r}+{\bf s})}{\epsilon_i+\epsilon_j}
\left( a_{ij}({\bf r}+{\bf s})+\frac{2{\rm g}n_{TF}({\bf r}+{\bf s})}
{\epsilon_i+\epsilon_j} b_{ij}({\bf r}+{\bf s}) \right) - 2n_{TF}({\bf r}+{\bf s})
b_{ij}({\bf r}+{\bf s})\delta\varphi({\bf r}+{\bf s}) \right] \;\;,
\nonumber
\end{eqnarray}
and
\begin{eqnarray}
i\hbar\omega\; \delta\varphi({\bf r}) &=& {\rm g} \left( 1+
{\rm g}\frac{1}{V}\sum_{\bf p}\frac{m}{p^2} \right)
\delta n_0({\bf r})
\label{hydro1}\\
&-& \frac{{\rm g}^2}{V^2}\sum_{ij} \frac{a_{ij}({\bf r})}{\epsilon_i+\epsilon_j} 
\int d{\bf s}\; e^{i{\bf s}\cdot\nabla[\varphi_i({\bf r})+\varphi_j({\bf r})]}\;
\delta n_0({\bf r}+{\bf s}) \left( a_{ij}({\bf r}+{\bf s})+\frac{2{\rm g}n_{TF}({\bf r}+{\bf s})}
{\epsilon_i+\epsilon_j} b_{ij}({\bf r}+{\bf s}) \right) \;\;,
\nonumber
\end{eqnarray}
where the matrices $a_{ij}$ and $b_{ij}$ have been defined in (\ref{ab}). By using the gradient 
expansion employed in the previous section for the calculation of the non-resonant contributions 
to  $\delta E$, we get the result  
\begin{eqnarray}
i\omega \left(1+\frac{4}{\sqrt{\pi}}(a^3n_{TF}({\bf r}))^{1/2}\right)\delta n_0({\bf r})
&=& \nabla \cdot\left[ n_0({\bf r})\left(1+\frac{8}{3\sqrt{\pi}}(a^3n_{TF}({\bf r}))^{1/2}\right)
{\bf v}_s({\bf r}) \right] \;\;,
\label{hydro2}
\end{eqnarray}
and
\begin{eqnarray}
i\hbar\omega\delta\varphi&=&{\rm g}\left(1+\frac{20}{\sqrt{\pi}}(a^2n_{TF})^{1/2}\right)
\delta n_0({\bf r}) \;\;.
\label{hydro3}
\end{eqnarray}
If one takes into account the effect of quantum depletion, the local relation between condensate 
density and total density is given by $n=n_0[1+8(a^3n_{TF})^{1/2}/(3\sqrt{\pi})]$, and for the  
fluctuations of the two densities $\delta n=\delta n_0[1+4(a^3n_{TF})^{1/2}/\sqrt{\pi}]$.
Finally, the change in the local chemical potential $\mu_l={\rm g}n[1+32(a^3n_{TF})^{1/2}/
(3\sqrt{\pi})]$ induced by a density fluctuation is given by the following expression
$\delta n\,\partial\mu_l/\partial n = \delta n_0 {\rm g}[1+20(a^3n_{TF})^{1/2}/\sqrt{\pi}]$.
It is now straightforward to recognize Eqs. (\ref{hydro2}), (\ref{hydro3}) as the linearized 
hydrodynamic equations 
\begin{eqnarray}
&& \frac{\partial \delta n}{\partial t} + \nabla\cdot\left(n{\bf v}_s\right) = 0 \;\;,
\nonumber\\
&& m\frac{\partial {\bf v}_s}{\partial t} + \nabla\cdot\left(\frac{\partial\mu_l}{\partial n}
\delta n\right) = 0 \;\;,
\label{hydro4}
\end{eqnarray}
which involve the total density $n$ and the superlfuid velocity ${\bf v}_s$.

\section{Concluding remarks}

In this paper we have studied the collisionless collective modes of a dilute Bose gas beyond the 
Gross-Pitaevskii theory. In particular, for harmonically trapped systems in the thermodynamic 
limit, we have calculated the corrections to the excitation frequencies of the low-lying collective 
modes. We find that, not far below the Bose-Einstein transition temperature, the fractional 
frequency shift is of the order of few percent for typical experimental conditions and 
can be measured. A direct comparison with experimental data obtained by the 
MIT group with large condensates looks very good. Similarly to what happened to 
Gross-Pitaevskii theory, the study of collective excitations can become a useful bench-mark 
also for theories beyond mean-field approximation.

\section*{Acknowledgments}

I would like to thank S. Stringari and L. Pitaevskii for many stimulating discussions. 
It is also a pleasure to thank A. Chikkatur for providing me with the experimental data 
of the frequency of the $m=0$ mode and D. Stamper-Kurn for useful remarks concerning these 
experimental results. Useful discussions with P. Fedichev, G. Shlyapnikov, A. Minguzzi 
and P. Schuck are also gratefully acknowledged. I am also particularly indebted to 
L. Pitaevskii for a critical reading of the manuscript.

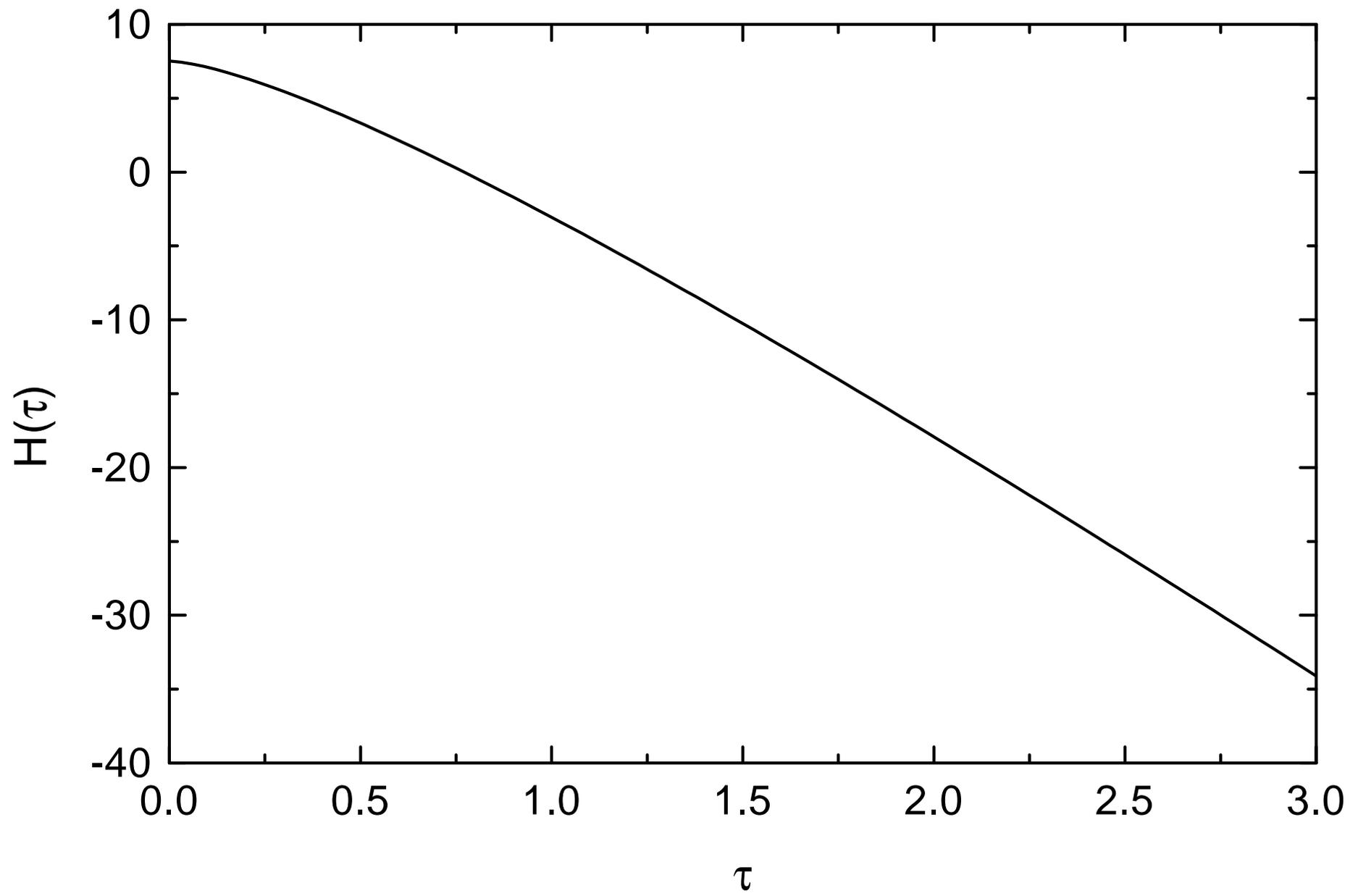
\begin{figure}
\caption{Dimensionless function $H$ as a function of the reduced temperature $\tau=k_BT/{\rm g}n_0$.}
\end{figure}

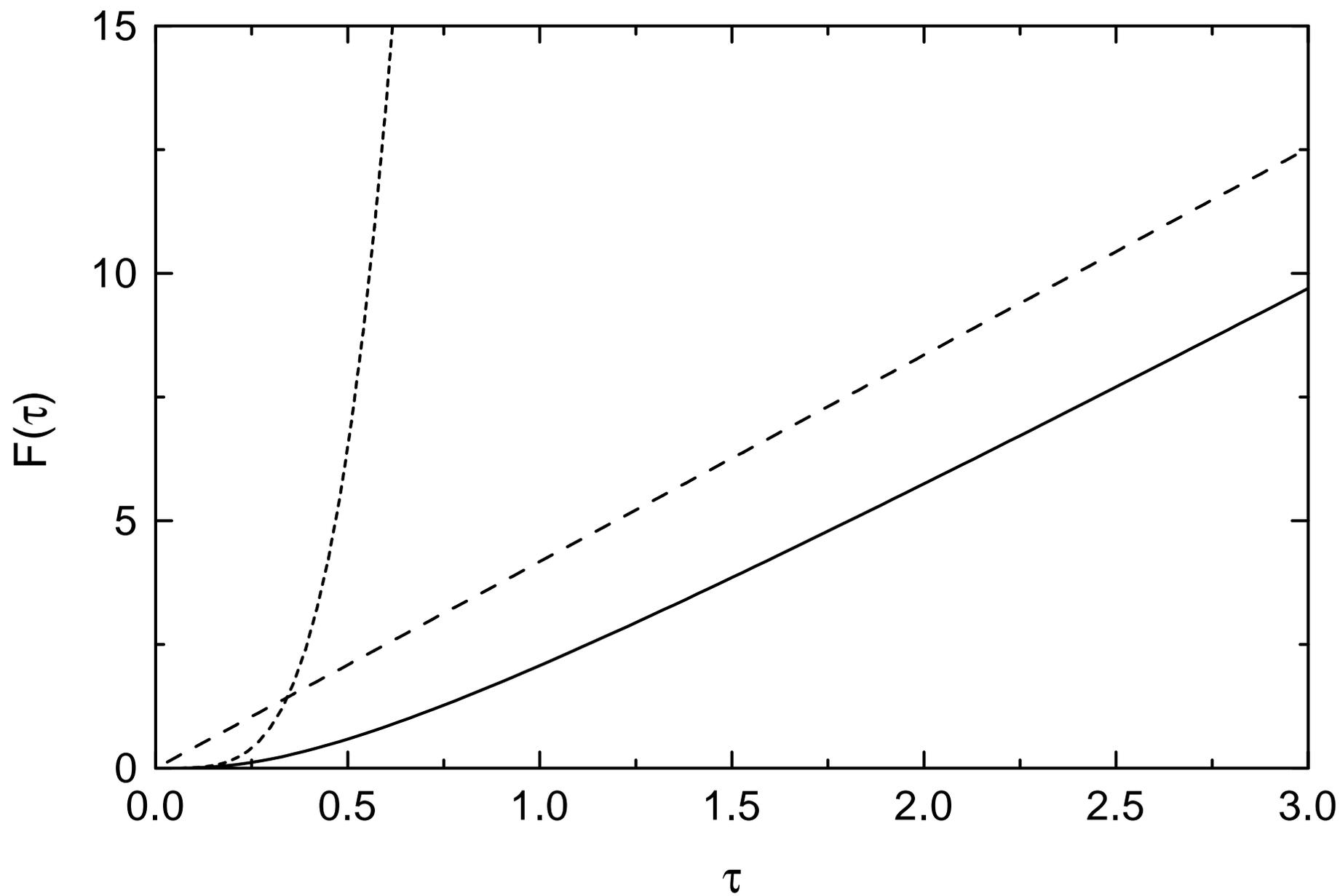
\begin{figure}
\caption{Dimensionless function $F$ as a function of the reduced temperature $\tau=k_BT/{\rm g}n_0$
(solid line). The asymptotic behaviors for $\tau\ll 1$ (dashed line) and for $\tau\gg 1$ (long-dashed 
line) are also reported.}
\end{figure}

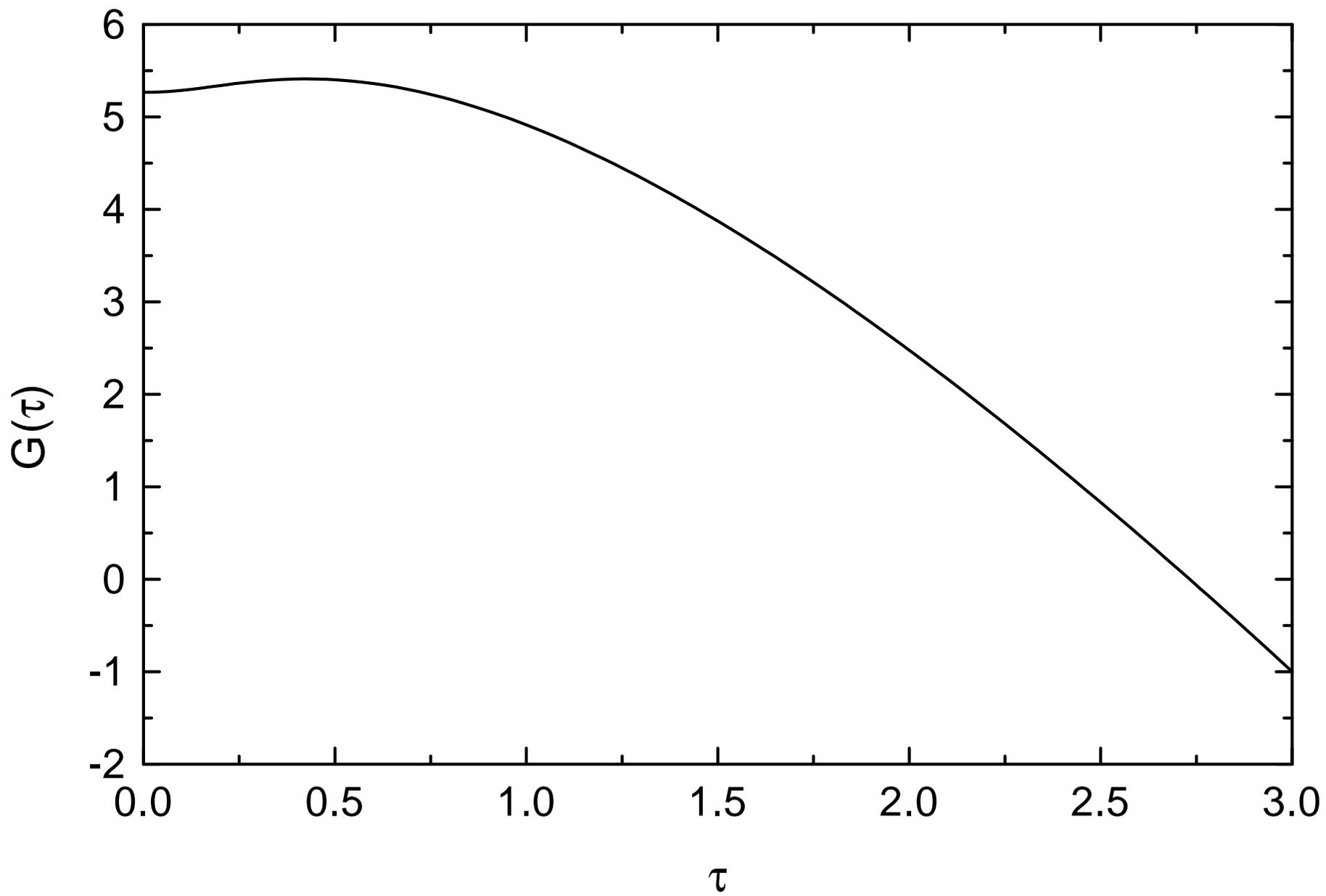
\begin{figure}
\caption{Dimensionless function $G$ as a function of the reduced temperature $\tau=k_BT/{\rm g}n_0$.}
\end{figure}

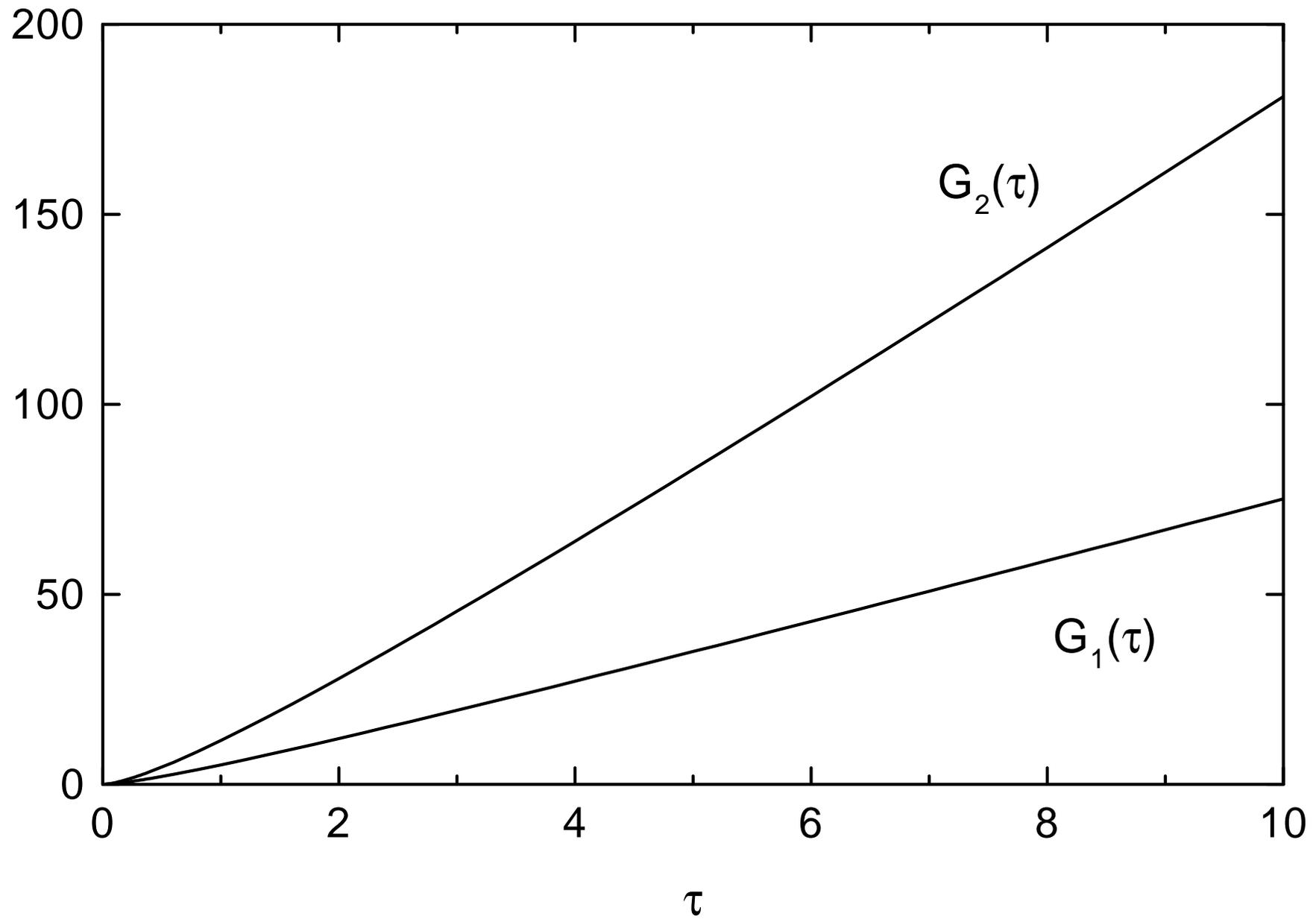
\begin{figure}
\caption{Dimensionless functions $G1$ and $G2$ as a function of the reduced temperature 
$\tau=k_BT/{\rm g}n_0$.}
\end{figure}

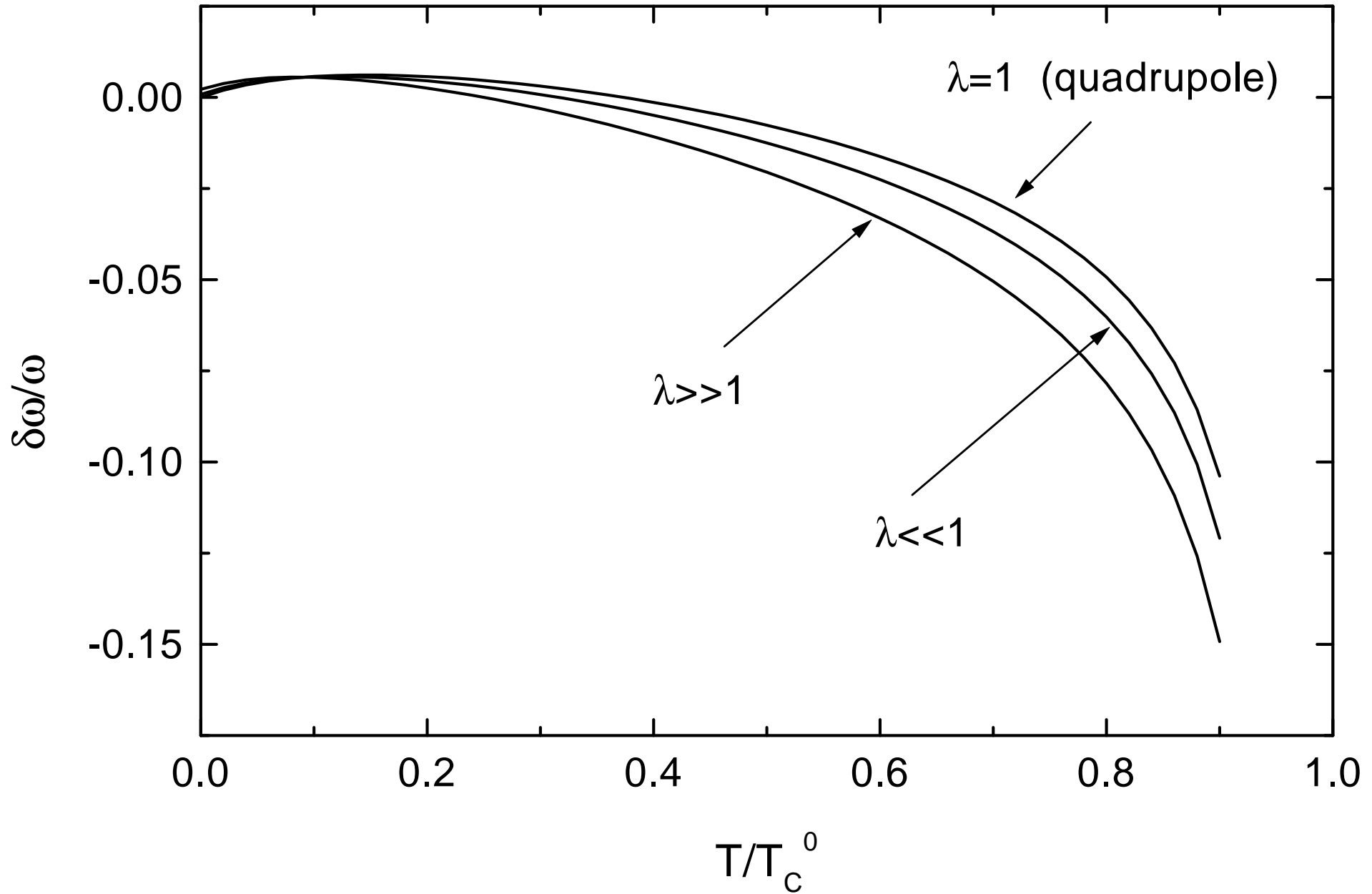
\begin{figure}
\caption{Fractional shift of the $m=0$ low mode as a function of the reduced temperature $T/T_c^0$.
The value of the interaction parameter is $\eta=0.4$. For a spherical trap ($\lambda=1$) this mode 
corresponds to the $l=2$ quadrupole mode.}
\end{figure}

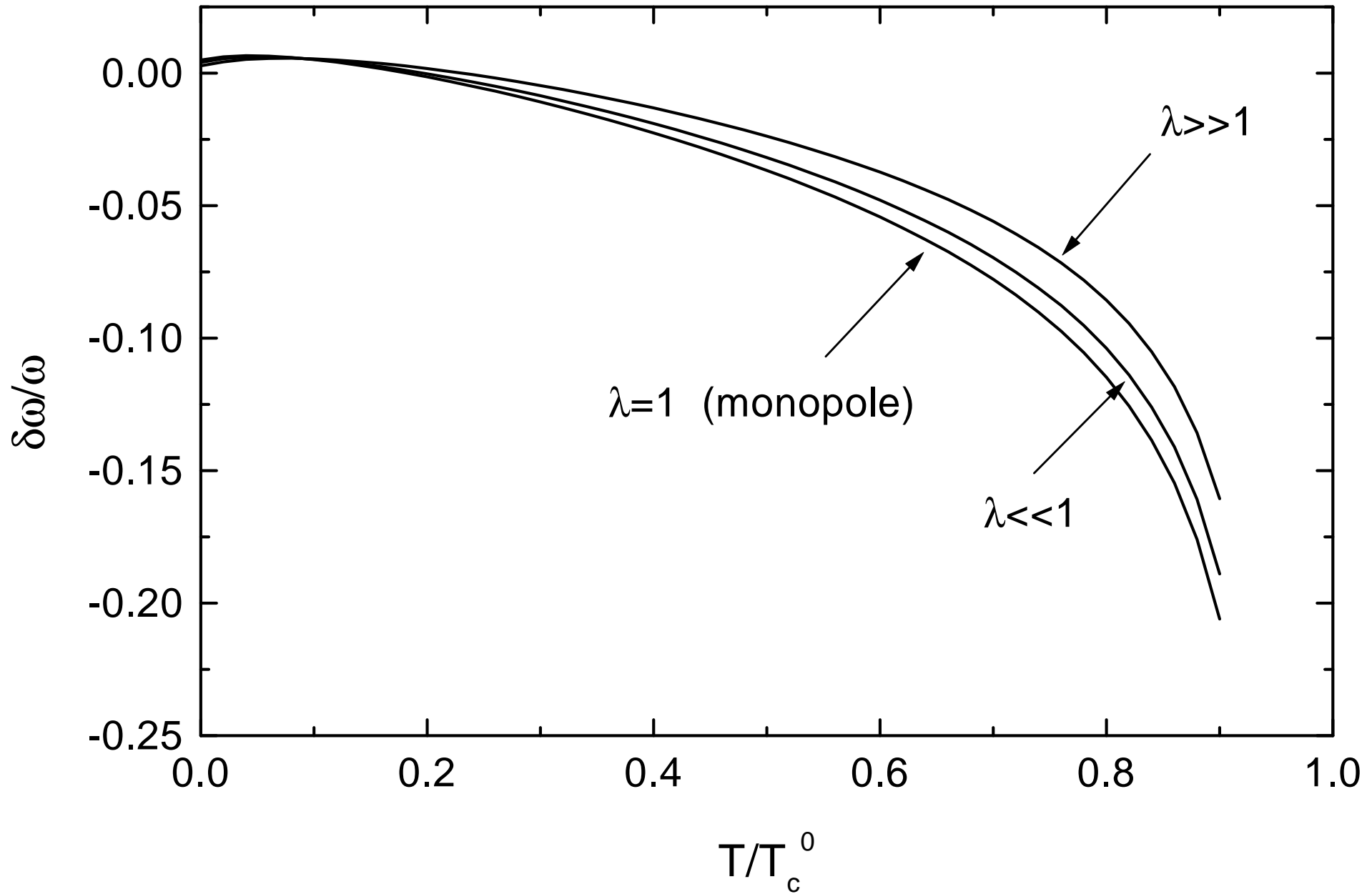
\begin{figure}
\caption{Fractional shift of the $m=0$ high mode as a function of the reduced temperature $T/T_c^0$.
The value of the interaction parameter is $\eta=0.4$. For a spherical trap ($\lambda=1$) this mode 
corresponds to the monopole (breathing) mode.}
\end{figure}

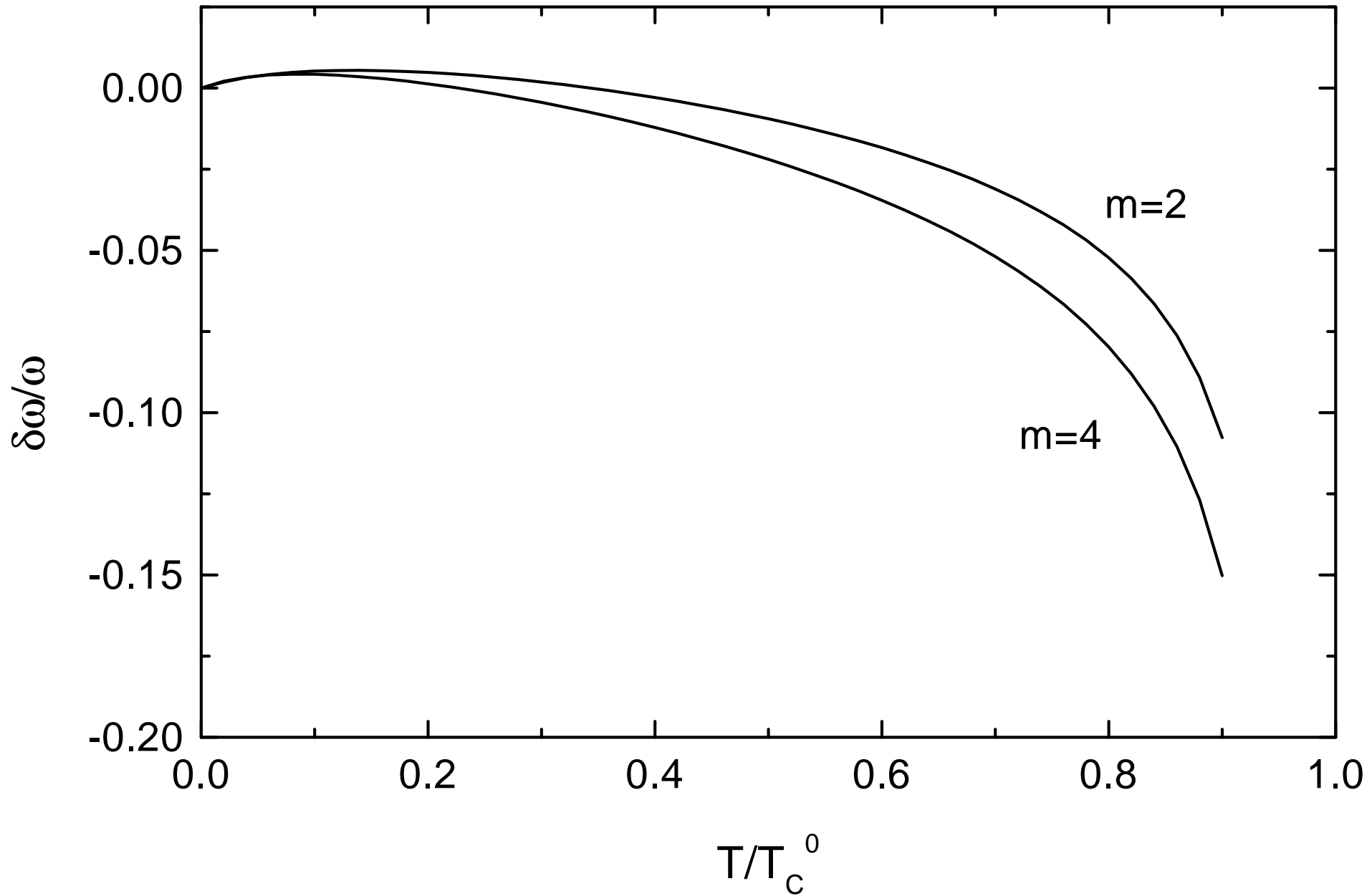
\begin{figure}
\caption{Fractional shift of the $m=2$ and $m=4$ surface modes as a function of the reduced 
temperature $T/T_c^0$. The value of the interaction parameter is $\eta=0.4$.}
\end{figure}

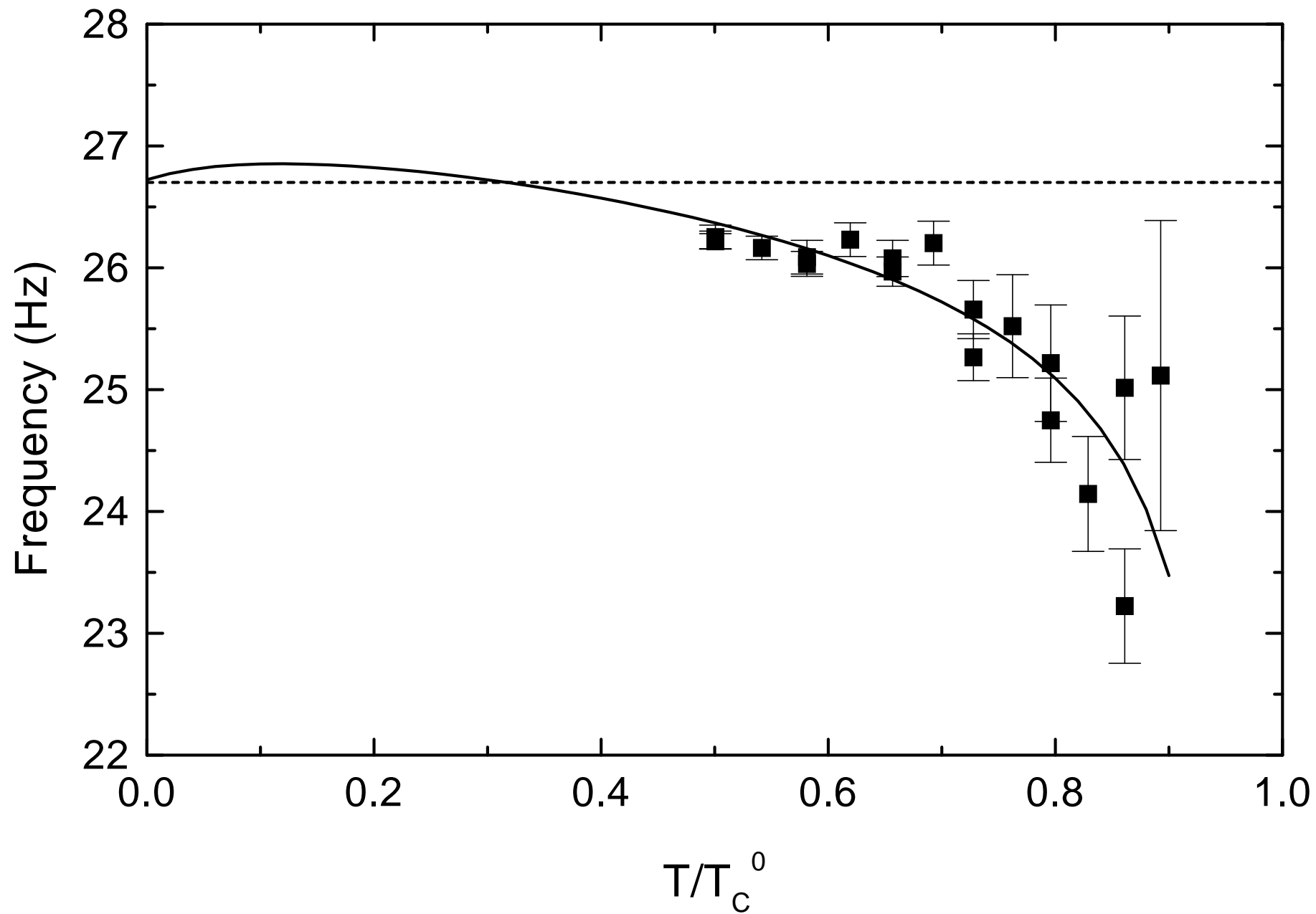
\begin{figure}
\caption{Temperature dependence of the frequency of the $m=0$ low mode. The experimental points 
are taken from \protect\cite{MIT98}. The theoretical calculation (solid line) corresponds to 
cigar-shaped traps and to the value $\eta=0.4$ of the interaction parameter. The dashed line 
corresponds to the hydrodynamic prediction $\nu=\protect\sqrt{5/2}\,\nu_z$ from 
Eq. (\protect\ref{m0LH}).}
\end{figure}


\begin{references}
\bibitem{N1} Not considered here are stationary states of the condensate containing
currents, such as vortex states.
\bibitem{N2} Time-dependent equations for $f_{ij}$ and $g_{ij}$ have been already 
derived in \cite{G98} within the Popov approximation. Here we include also 
the coupling to $\delta\tilde{n}$ and $\delta\tilde{m}$, which was neglected in 
\cite{G98}. The reader is referred to \cite{G98} for the details of the 
calculation.
\bibitem{DSK} The measurements of the number of trapped atoms from phase-contrast images in 
\cite{MIT98} are about 25\% too large at the transition temperature (D. Stamper-Kurn, private
communication). The reduced temperature $T/T_c^0$ in Fig. 8 has been determined taking into 
account this normalization factor.
\end{references}
\end{document}